\documentclass[10pt]{article}
\usepackage{graphicx}
\usepackage{amsmath}
\usepackage{amssymb}
\usepackage{caption2}
\begin{document}
\bibliographystyle{prsty}
\begin{center}
{\large {\bf \sc{  Reanalysis of  the $Y(3940)$, $Y(4140)$, $Z_c(4020)$, $Z_c(4025)$ and $Z_b(10650)$ as   molecular states
with  QCD sum rules }}} \\[2mm]
Zhi-Gang  Wang \footnote{E-mail: zgwang@aliyun.com.  }     \\
 Department of Physics, North China Electric Power University, Baoding 071003, P. R. China
\end{center}

\begin{abstract}
In this article, we calculate the contributions of the vacuum condensates up to dimension-10 in the operator product expansion, and  study the
$J^{PC}=0^{++}$, $1^{+-}$, $2^{++}$ $D^*\bar{D}^*$, $D_s^*\bar{D}_s^*$,  $B^*\bar{B}^*$, $B_s^*\bar{B}_s^*$ molecular states   with the QCD sum rules. In calculations,  we use the  formula $\mu=\sqrt{M^2_{X/Y/Z}-(2{\mathbb{M}}_Q)^2}$  to determine  the energy scales of the QCD spectral densities.  The numerical results favor assigning the $Z_c(4020)$ and $Z_c(4025)$  as the $J^{PC}=0^{++}$, $1^{+-}$ or $2^{++}$   $D^*\bar{D}^*$ molecular states, the $Y(4140)$ as the $J^{PC}=0^{++}$ $D^*_s{D}_s^*$ molecular state, the $Z_b(10650)$  as the $J^{PC}=1^{+-}$    $B^*\bar{B}^*$ molecular state, and  disfavor assigning the $Y(3940)$  as the ($J^{PC}=0^{++}$) molecular state.  The present  predictions can be confronted with  the experimental data in the futures.
\end{abstract}



\section{Introduction}

In 2004, the  Belle collaboration  observed  the near-threshold enhancement $Y(3940)$  in the $\omega J/\psi$  mass spectrum in the
 exclusive $B \to K \omega J/\psi$ decays \cite{Belle2004}. In 2007, the BaBar collaboration confirmed the $Y(3940)$ in the  exclusive $B \to K \omega J/\psi$ decays \cite{BaBar2007}. In 2010, the   Belle collaboration confirmed the $Y(3940)$    in the process $\gamma\gamma\to \omega J/\psi$ \cite{Belle2010}. Now the $X(3915)$ ($Y(3940)$) is listed in the Review of Particle Physics as the $\chi_{c0}({\rm 2P})$
state with the quantum numbers $J^{PC}=0^{++}$ \cite{PDG}.

In 2009, the CDF collaboration observed the narrow structure $Y(4140)$ near the $J/\psi\phi$ threshold in the exclusive $B^+\to J/\psi\phi K^+$ decays \cite{CDF0903}. Latter, the Belle collaboration  searched for  the $Y(4140)$ in the process $\gamma \gamma \to \phi J/\psi$ and observed no evidence \cite{Belle0912}. In 2012, the LHCb collaboration  searched  for the $Y(4140)$ state in $B^+ \to J/\psi \phi K^+$ decays, and observed no evidence \cite{LHCb1202}.
In 2013, the CMS collaboration observed a peaking structure consistent with the $Y(4140)$ in the $J/\psi\phi$  mass spectrum in the $B^\pm \to J/\psi \phi K^\pm$ decays,  and fitted the structure to a $S$-wave relativistic Breit-Wigner line-shape with the statistical significance exceeding $5 \sigma$ \cite{CMS1309}.
Also in 2013,  the D0 collaboration observed the  $Y(4140)$  in the  $B^+ \to J/\psi \phi K^+$ decays  with the statistical significance of $3.1\sigma$  \cite{D0-1309}.  However, there is no suitable position in the $\bar{c}c$ spectroscopy for the $Y(4140)$.

The $Y(3940)$ and $Y(4140)$ appear  near  the $D^*\bar{D}^*$ and $D_s^*\bar{D}_s^*$ thresholds respectively, and  have analogous decays,
\begin{eqnarray}
Y(3940)&\to&J/\psi\,\varphi \, , \nonumber\\
Y(4140)&\to&J/\psi\,\phi \, .
\end{eqnarray}
It is natural to relate  the $Y(3940)$ and $Y(4140)$ with the $D^*\bar{D}^*$ and $D_s^*\bar{D}_s^*$ molecular states, respectively \cite{Y4140-molecule,Y4140-Wang,Y4140-Nielsen,Y4140-Zhang}. Other assignments, such as the hybrid charmonium states \cite{Y4140-Wang,Y4140-hybrid} and tetraquark states \cite{Y4140-tetraquark} also suggested.

In 2011, the Belle collaboration  observed  the $Z_b(10610)$ and $Z_b(10650)$ in the $\pi^{\pm}\Upsilon({\rm 1,2,3S})$  and $\pi^{\pm} h_b({\rm 1,2P})$  invariant mass distributions  in the $\Upsilon({\rm 5S})\to \pi^{+}\pi^-\Upsilon({\rm 1,2,3S})$, $\pi^{+}\pi^- h_b({\rm 1,2P})$ decays \cite{Belle1105}.  The quantum numbers   $I^G(J^P)=1^+(1^+)$ are favored \cite{Belle1105}.
Later, the Belle collaboration updated the measured parameters  $ M_{Z_b(10610)}=(10607.2\pm2.0)\,\rm{ MeV}$, $M_{Z_b(10650)}=(10652.2\pm1.5)\,\rm{MeV}$, $\Gamma_{Z_b(10610)}=(18.4\pm2.4) \,\rm{MeV}$ and
$\Gamma_{Z_b(10650)}=(11.5\pm2.2)\,\rm{ MeV}$ \cite{Belle1110}.
In 2013, the Belle collaboration observed  the  $Z_b^0(10610)$ in a Dalitz
analysis of the decays to $\Upsilon(2,3{\rm S}) \pi^0$
in the $\Upsilon(5{\rm S}) \to \Upsilon ({\rm 1,2,3S}) \pi^0 \pi^0$ decays
\cite{Belle1308}.
The $Z_b(10610)$ and $Z_b(10650)$  appear  near the $B\bar{B}^*$ and $B^*\bar{B}^*$ thresholds, respectively.
It is natural to relate  the $Z_b(10610)$ and $Z_b(10650)$ with the  $B\bar{B}^*$ and $B^*\bar{B}^*$ molecular states, respectively \cite{Molecule-Zb,WangHuang-molecule}. Other assignments, such as the  tetraquark states \cite{Tetraquark-Zb,Tetraquark-Zb-QCDSR,WangHuangTao1312}, threshold cusps \cite{Cusp-Zb}, the re-scattering effects \cite{Rescatter-Zb}, etc are also suggested.

 In 2013, the BESIII collaboration  observed
the $Z^{\pm}_c(4025)$ near the $(D^{*} \bar{D}^{*})^{\pm}$ threshold in the $\pi^\mp$ recoil mass spectrum  in the process $e^+e^- \to (D^{*} \bar{D}^{*})^{\pm} \pi^\mp$ \cite{BES1308}. Furthermore, the  BESIII collaboration observed the  $Z_c(4020)$   in the $\pi^\pm h_c$ mass spectrum in the process $e^+e^- \to \pi^+\pi^- h_c$ \cite{BES1309}.
The $Z_c(4020)$ and $Z_c(4025)$  appear  near the $D^*\bar{D}^*$ threshold. It is natural to relate them with the $D^*\bar{D}^*$ molecular states \cite{Molecule,Cui-DvDv,Chen-Zhu,DvDv-Nielsen}. Other assignments, such as the re-scattering effects \cite{Rescatter},  tetraquark states \cite{Tetraquark-Qiao,Wang1311}, etc are also suggested.

The $Z_c(4020)$, $Z_c(4025)$, $Z_b(10610)$, $Z_b(10650)$ appear  near  the $D^*\bar{D}^*$, $D^*\bar{D}^*$,  $B\bar{B}^*$, $B^*\bar{B}^*$ thresholds respectively, and have  analogous decays
\begin{eqnarray}
Z_c^{\pm}(4020) &\to& \pi^\pm \, h_c \, , \nonumber\\
Z_c^{\pm}(4025) &\to& (D^*\bar{D}^*)^\pm  \, , \nonumber\\
Z_b^{\pm}(10610) &\to& \pi^{\pm}\Upsilon({\rm 1,2,3S})\, ,  \, \pi^{\pm} h_b({\rm 1,2P})   \, , \nonumber\\
Z_b^{\pm}(10650) &\to& \pi^{\pm}\Upsilon({\rm 1,2,3S})\, ,  \, \pi^{\pm} h_b({\rm 1,2P})   \, .
\end{eqnarray}
The $S$-wave $D^{*} \bar{D}^{*}$, $D_s^{*} \bar{D}_s^{*}$, $B^{*} \bar{B}^{*}$, $B_s^{*} \bar{B}_s^{*}$ systems  have the quantum numbers $J^{PC}=0^{++}$, $1^{+-}$, $2^{++}$,  the $S$-wave $ \pi^\pm h_Q$ systems have the quantum numbers $J^{PC}=1^{--}$, the $S$-wave $ \pi^\pm \Upsilon$ systems have the quantum numbers $J^{PC}=1^{+-}$.
 It is also possible for the $P$-wave $\pi^{\pm} h_Q$ systems to have the quantum numbers $J^{PC}=0^{++}$, $1^{+-}$, $2^{++}$.

In this article, we  take the $Y(3940)$, $Z_c(4020)$, $Z_c(4025)$ as the $D^{*} \bar{D}^{*}$ molecular states, the $Y(4140)$ as the $D_s^{*} \bar{D}_s^{*}$ molecular state, the $Z_b(10610)$ as the $B\bar{B}^*$ molecular state, the $Z_b(10650)$ as the $B^*\bar{B}^*$ molecular state, study the $J^{PC}=0^{++}$, $1^{+-}$, $2^{++}$ molecular states consist of $D^*\bar{D}^*$, $D_s^*\bar{D}_s^*$,  $B^*\bar{B}^*$, $B_s^*\bar{B}_s^*$ with the QCD sum rules,  and make tentative  assignments of the $Y(3940)$, $Y(4140)$, $Z_c(4020)$, $Z_c(4025)$ and $Z_b(10650)$ in the scenario of molecular states.

In Ref.\cite{Y4140-Wang}, we study the scalar $D^*\bar{D}^*$,  $D_s^*\bar{D}_s^*$, $B^*\bar{B}^*$,  $B_s^*\bar{B}_s^*$ molecular states with the QCD sum rules by  carrying  out the operator product expansion to the vacuum condensates  up to
dimension-10 and setting the energy scale to be $\mu=1\,\rm{GeV}$. The predicted masses
 disfavor assigning the $Y(4140)$ as the scalar
${D}_s^\ast {\bar {D}}_s^\ast$ molecular state.
In Ref.\cite{Y4140-Nielsen}(\cite{Y4140-Zhang}), R. M. Albuquerque et al (Zhang and Huang) study the scalar   $D_s^*\bar{D}_s^*$ molecular state with the QCD sum rules by carrying   out the operator product expansion to the vacuum condensates up to dimension-8 (6), and their predictions  favor assigning the $Y(4140)$ as the $J^P=0^+$ molecular state,  but they do not show or do not specify  the energy scales of the QCD spectral densities.
In Refs.\cite{Tetraquark-Zb-QCDSR,Cui-DvDv},  Cui, Liu and Huang  study the  axial-vector  $B^*\bar{B}^*$ ($D^*\bar{D}^*$) molecular state with the QCD sum rules by carrying   out the operator product expansion to the vacuum condensates up to dimension-6, and their predictions favor assigning the $Z_b(10650)$ ($Z_c(4025)$) as the axial-vector   $B^*\bar{B}^*$ ($D^*\bar{D}^*$) molecular state,  but they do not show or do not specify  the energy scales of the QCD spectral densities.
Furthermore, in Refs.\cite{Y4140-Nielsen,Y4140-Zhang,Tetraquark-Zb-QCDSR,Cui-DvDv},  some higher dimension vacuum condensates involving the gluon condensate, mixed condensate and four-quark condensate are neglected, which  impairs the predictive ability, as the higher dimension vacuum condensates play an important role in determining the Borel windows.

In this article, we study the $J^{PC}=0^{++}$, $1^{+-}$, $2^{++}$ molecular states consist of $D^*\bar{D}^*$, $D_s^*\bar{D}_s^*$,  $B^*\bar{B}^*$, $B_s^*\bar{B}_s^*$ with the QCD sum rules according to the routine in our previous works \cite{WangHuang-molecule,WangHuangTao1312,Wang1311,WangHuangTao}.

In Refs.\cite{WangHuangTao1312,Wang1311,WangHuangTao}, we focus on the scenario of tetraquark states, calculate the  vacuum condensates up to dimension-10  in
the operator product expansion, study the diquark-antidiquark type scalar, vector, axial-vector, tensor hidden charmed tetraquark states and
axial-vector hidden bottom tetraquark states systematically  with the QCD sum rules, and make reasonable  assignments of the $X(3872)$,
$Z_c(3900)$, $Z_c(3885)$, $Z_c(4020)$, $Z_c(4025)$, $Z(4050)$, $Z(4250)$, $Y(4360)$, $Y(4630)$, $Y(4660)$, $Z_b(10610)$  and $Z_b(10650)$.
In Ref.\cite{WangHuang-molecule}, we focus on the scenario of molecular states,
 calculate the  vacuum condensates up to dimension-10  in
the operator product expansion, study the  axial-vector  hadronic molecular states with the QCD sum rules, and make tentative assignments  of the $X(3872)$, $Z_c(3900)$, $Z_b(10610)$. The interested reader can consult Ref.\cite{Swanson2006} for more articles  on the exotic  $X$, $Y$ and $Z$ particles. A hadron cannot be identified  unambiguously by  the mass alone. It is interesting to explore possible assignments in the scenario of molecular states.

  In Refs.\cite{WangHuang-molecule,WangHuangTao1312,Wang1311,WangHuangTao}, we    explore the energy scale dependence of the hidden charmed (bottom) tetraquark states and molecular states in details for the first time, and suggest a  formula
\begin{eqnarray}
\mu&=&\sqrt{M^2_{X/Y/Z}-(2{\mathbb{M}}_Q)^2} \, ,
 \end{eqnarray}
 with the effective masses ${\mathbb{M}}_Q$ to determine  the energy scales of the  QCD spectral densities in the QCD sum rules, which works very well.

 In this article, we  calculate the contributions of the vacuum condensates up to dimension-10  in a  consistent way, study the $J^{PC}=0^{++}$, $1^{+-}$, $2^{++}$ molecular states consist of $D^*\bar{D}^*$, $D_s^*\bar{D}_s^*$,  $B^*\bar{B}^*$, $B_s^*\bar{B}_s^*$  in a systematic way, and make tentative  assignments of the $Y(3940)$, $Y(4140)$, $Z_c(4020)$, $Z_c(4025)$ and $Z_b(10650)$ based on the QCD sum rules.

The article is arranged as follows:  we derive the QCD sum rules for the masses and pole residues of  the $D^*\bar{D}^*$, $D_s^*\bar{D}_s^*$,  $B^*\bar{B}^*$, $B_s^*\bar{B}_s^*$ molecular  states  in section 2; in section 3, we present the numerical results and discussions; section 4 is reserved for our conclusion.

\section{QCD sum rules for  the  $D^*\bar{D}^*$, $D_s^*\bar{D}_s^*$,  $B^*\bar{B}^*$, $B_s^*\bar{B}_s^*$ molecular states }
In the following, we write down  the two-point correlation functions $\Pi_{\mu\nu\alpha\beta}(p)$ and $\Pi(p)$ in the QCD sum rules,
\begin{eqnarray}
\Pi_{\mu\nu\alpha\beta}(p)&=&i\int d^4x e^{ip \cdot x} \langle0|T\left\{\eta_{\mu\nu}(x)\eta_{\alpha\beta}^{\dagger}(0)\right\}|0\rangle \, ,\\
\Pi(p)&=&i\int d^4x e^{ip \cdot x} \langle0|T\left\{\eta(x)\eta^{\dagger}(0)\right\}|0\rangle \, ,
\end{eqnarray}

\begin{eqnarray}
 \eta^{\pm}_{\bar{u}d;\mu\nu}(x)&=&\frac{\bar{u}(x)\gamma_\mu Q(x) \bar{Q}(x)\gamma_\nu d(x)\pm\bar{u}(x)\gamma_\nu Q(x) \bar{Q}(x)\gamma_\mu d(x)}{\sqrt{2}} \, , \nonumber \\
  \eta^{\pm}_{\bar{s}s;\mu\nu}(x)&=&\frac{\bar{s}(x)\gamma_\mu Q(x) \bar{Q}(x)\gamma_\nu s(x)\pm\bar{s}(x)\gamma_\nu Q(x) \bar{Q}(x)\gamma_\mu s(x)}{\sqrt{2}} \, , \nonumber \\
  \eta_{\bar{u}d}(x)&=&\bar{u}(x)\gamma_\mu Q(x) \bar{Q}(x)\gamma^\mu d(x)\, ,   \nonumber\\
   \eta_{\bar{s}s}(x)&=&\bar{s}(x)\gamma_\mu Q(x) \bar{Q}(x)\gamma^\mu s(x)\, ,
\end{eqnarray}
where $\eta_{\mu\nu}(x)=\eta^{\pm}_{\bar{u}d;\mu\nu}(x),\,\eta^{\pm}_{\bar{s}s;\mu\nu}(x)$, $\eta(x)=\eta^{\pm}_{\bar{u}d}(x),\,\eta^{\pm}_{\bar{s}s}(x)$, $Q=c,b$.
 Under charge conjugation transform $\widehat{C}$, the currents $\eta^{\pm}_{\mu\nu}(x)$ and $\eta(x)$ have the following properties,
\begin{eqnarray}
\widehat{C}\,\eta^{\pm}_{\mu\nu}(x)\,\widehat{C}^{-1}&=&\pm \,\eta_{\mu\nu}^{\pm}(x)\mid_{u \leftrightarrow d}\, , \nonumber \\
\widehat{C}\,\eta(x)\,\widehat{C}^{-1}&=& \eta(x)\mid_{u \leftrightarrow d} \, ,
\end{eqnarray}
thereafter we will smear the subscripts $\bar{u}d$, $\bar{s}s$ and superscripts $\pm$ for simplicity. On the other hand, the  currents $\eta^{\pm}_{\mu\nu}(x)$ and $\eta(x)$  are of the type $V_\mu{\otimes}V_\nu $, where the $V_\mu$ denotes  the two-quark vector currents interpolating the conventional vector heavy mesons, so they have positive parity.
The currents $\eta^+_{\mu\nu}(x)$ and $\eta(x)$ have both positive charge conjugation and positive parity, therefore  couple potentially to the $J^{PC}=2^{++}$ or $0^{++}$ states, while the currents $\eta^{-}_{\mu\nu}(x)$ have negative charge conjugation but positive parity, therefore  couple potentially to the $J^{PC}=1^{+-}$ states.
We construct the color singlet-singlet type currents $\eta_{\mu\nu}(x)$ and $\eta(x)$ to study the   $D^*\bar{D}^*$, $D_s^*\bar{D}_s^*$,  $B^*\bar{B}^*$, $B_s^*\bar{B}_s^*$ molecular states, and assume that the operators  $\eta_{\mu\nu}(x)$ and $\eta(x)$ couple potentially to the bound states, not to the scattering states.
We can also construct the color octet-octet type currents $\eta^8_{\mu\nu}(x)$ and $\eta^8(x)$, which have the same quantum numbers $J^{PC}$ as their color singlet-singlet partners,  to study the   $D^*\bar{D}^*$, $D_s^*\bar{D}_s^*$,  $B^*\bar{B}^*$, $B_s^*\bar{B}_s^*$ molecular states,
\begin{eqnarray}
 \eta^{8\pm}_{\bar{u}d;\mu\nu}(x)&=&\frac{\bar{u}(x)\gamma_\mu \lambda^a Q(x) \bar{Q}(x)\gamma_\nu \lambda^a d(x)\pm\bar{u}(x)\gamma_\nu \lambda^a Q(x) \bar{Q}(x)\gamma_\mu \lambda^a d(x)}{\sqrt{2}} \, , \nonumber \\
  \eta^{8\pm}_{\bar{s}s;\mu\nu}(x)&=&\frac{\bar{s}(x)\gamma_\mu \lambda^a Q(x) \bar{Q}(x)\gamma_\nu \lambda^a s(x)\pm\bar{s}(x)\gamma_\nu \lambda^aQ(x) \bar{Q}(x)\gamma_\mu \lambda^as(x)}{\sqrt{2}} \, , \nonumber \\
  \eta^8_{\bar{u}d}(x)&=&\bar{u}(x)\gamma_\mu \lambda^a Q(x) \bar{Q}(x)\gamma^\mu \lambda^a d(x)\, ,   \nonumber\\
   \eta^8_{\bar{s}s}(x)&=&\bar{s}(x)\gamma_\mu \lambda^a Q(x) \bar{Q}(x)\gamma^\mu \lambda^a s(x)\, ,
\end{eqnarray}
where the $\lambda^a$ is the Gell-Mann matrix. In Ref.\cite{WangHuang-molecule}, we observe that the color octet-octet type
molecular states have larger masses than that of the corresponding color singlet-singlet type molecular states. So in this article, we prefer the color singlet-singlet type currents, which couple potentially to the color singlet-singlet type molecular states have smaller masses.
In Refs.\cite{Tetraquark-Zb-QCDSR,Cui-DvDv},  Cui, Liu and  Huang take the currents $j_\mu(x)$,
\begin{eqnarray}
j_\mu(x)&=&\epsilon_{\mu\nu\alpha\beta}\, \bar{u}(x)\gamma^{\nu}Q(x)\,iD^{\alpha}\,\bar{Q}(x)\gamma^{\beta}d(x)\, ,
\end{eqnarray}
where $D^\alpha=\partial^\alpha-ig_sG^\alpha(x)$, to study the  $Z_b(10650)$ and $Z_c(4025)$ as the $B^*\bar{B}^*$ and $D^*\bar{D}^*$ molecular states respectively with $J^{P}=1^+$.
In Ref.\cite{Chen-Zhu}, W. Chen et al take the current $J_\mu(x)$,
\begin{eqnarray}
J_\mu (x)&=& \bar{q}(x)\gamma^{\alpha}c(x) \bar{c}(x)\sigma_{\alpha\mu}\gamma_5 q(x)-\bar{q}(x)\sigma_{\alpha\mu}\gamma_5c(x) \bar{c}(x)\gamma^{\alpha} q(x)\, ,
\end{eqnarray}
to study the $Z_c(4025)$ as the  $D^*\bar{D}^*$ molecular state with $J^{PC}=1^{+-}$. In this article, we use the simple $V_\mu {\otimes} V_\nu$ type currents to study the $J^{PC}=0^{++},\,1^{+-},\,2^{++}$ molecular states in a systematic way.

At the hadronic side, we can insert  a complete set of intermediate hadronic states with
the same quantum numbers as the current operators $\eta_{\mu\nu}(x)$ and $\eta(x)$ into the
correlation functions $\Pi_{\mu\nu\alpha\beta}(p)$ and $\Pi(p)$ to obtain the hadronic representation
\cite{SVZ79,Reinders85}. After isolating the ground state
contributions of the scalar, axial-vector and tensor molecular states, we get the following results,
\begin{eqnarray}
\Pi_{\mu\nu\alpha\beta}^{J=2}(p)&=&\Pi_{J=2}(p)\left( \frac{\widetilde{g}_{\mu\alpha}\widetilde{g}_{\nu\beta}+\widetilde{g}_{\mu\beta}\widetilde{g}_{\nu\alpha}}{2}-\frac{\widetilde{g}_{\mu\nu}\widetilde{g}_{\alpha\beta}}{3}\right) +\Pi_s(p)\,g_{\mu\nu}g_{\alpha\beta}   \, , \nonumber\\
&=&\frac{\lambda_{Y/Z}^2}{M_{Y/Z}^2-p^2}\left( \frac{\widetilde{g}_{\mu\alpha}\widetilde{g}_{\nu\beta}+\widetilde{g}_{\mu\beta}\widetilde{g}_{\nu\alpha}}{2}-\frac{\widetilde{g}_{\mu\nu}\widetilde{g}_{\alpha\beta}}{3}\right) +\cdots \, \, , \\
\Pi_{\mu\nu\alpha\beta}^{J=1}(p)&=&\Pi_{J=1}(p)\left( -\widetilde{g}_{\mu\alpha}p_{\nu}p_{\beta}-\widetilde{g}_{\nu\beta}p_{\mu}p_{\alpha}+\widetilde{g}_{\mu\beta}p_{\nu}p_{\alpha}+\widetilde{g}_{\nu\alpha}p_{\mu}p_{\beta}\right) +\Pi_s(p)\left(g_{\mu\alpha}g_{\nu\beta}-g_{\mu\beta}g_{\nu\alpha}\right) \, \, , \nonumber\\
&=&\frac{\lambda_{ Y/Z}^2}{M_{Y/Z}^2-p^2}\left( -\widetilde{g}_{\mu\alpha}p_{\nu}p_{\beta}-\widetilde{g}_{\nu\beta}p_{\mu}p_{\alpha}+\widetilde{g}_{\mu\beta}p_{\nu}p_{\alpha}+\widetilde{g}_{\nu\alpha}p_{\mu}p_{\beta}\right) +\cdots \, \, , \\
\Pi^{J=0}(p)&=&\Pi_{J=0}(p)=\frac{\lambda_{ Y/Z}^2}{M_{Y/Z}^2-p^2} +\cdots \, \, ,
\end{eqnarray}
where the notation $\widetilde{g}_{\mu\nu}=g_{\mu\nu}-\frac{p_\mu p_\nu}{p^2}$, the components  $\Pi_s(p)$ are irrelevant in the present analysis \cite{WangHcHb},  and the pole residues  $\lambda_{Y/Z}$ are defined by
\begin{eqnarray}
 \langle 0|\eta^{+}_{\mu\nu}(0)|{Y/Z}_{J=2}(p)\rangle &=& \lambda_{Y/Z} \, \varepsilon_{\mu\nu} \, , \nonumber\\
 \langle 0|\eta^{-}_{\mu\nu}(0)|{Y/Z}_{J=1}(p)\rangle &=& \lambda_{Y/Z} \,\left(\varepsilon_{\mu}p_{\nu}-\varepsilon_{\nu}p_{\mu} \right)\, , \nonumber\\
  \langle 0|\eta(0)|{Y/Z}_{J=0}(p)\rangle &=& \lambda_{Y/Z} \, ,
\end{eqnarray}
the $\varepsilon_{\mu\nu}$ and $\varepsilon_\mu$ are the polarization vectors of the tensor and axial-vector molecular states respectively  with the following properties,
 \begin{eqnarray}
 \sum_{\lambda}\varepsilon^*_{\alpha\beta}(\lambda,p)\varepsilon_{\mu\nu}(\lambda,p)
 &=&\frac{\widetilde{g}_{\alpha\mu}\widetilde{g}_{\beta\nu}+\widetilde{g}_{\alpha\nu}\widetilde{g}_{\beta\mu}}{2}-\frac{\widetilde{g}_{\alpha\beta}\widetilde{g}_{\mu\nu}}{3}\, , \nonumber \\
\sum_{\lambda}\varepsilon^*_{\mu}(\lambda,p)\varepsilon_{\nu}(\lambda,p)&=&-\widetilde{g}_{\mu\nu} \, .
 \end{eqnarray}
Here we add the superscripts and subscripts $J=2,\,1,\,0$ to denote the total angular momentum.
 In Ref.\cite{DvDv-Nielsen}, K. P. Khemchandani et al take the current
$j_{\mu\nu}(x)=\bar{c}(x)\gamma_\mu u(x) \bar{d}(x)\gamma_\nu c(x)$
to interpolate the molecular states, and use the  projectors
${\mathcal{P}^0}= \frac{\widetilde{g}_{\mu\nu}\widetilde{g}_{\alpha\beta}}{3}$, ${\mathcal{P}^1}= \frac{\widetilde{g}_{\mu\alpha}\widetilde{g}_{\nu\beta}-\widetilde{g}_{\mu\beta}\widetilde{g}_{\nu\alpha}}{2}$,
${\mathcal{P}^2}= \frac{\widetilde{g}_{\mu\alpha}\widetilde{g}_{\nu\beta}+\widetilde{g}_{\mu\beta}\widetilde{g}_{\nu\alpha}}{2}-\frac{\widetilde{g}_{\mu\nu}\widetilde{g}_{\alpha\beta}}{3}\, $ to separate the contributions of the  $J^P=0^+,\,1^+,\,2^+$ molecular states, respectively. The present treatment differs  from  that of Ref.\cite{DvDv-Nielsen}, while the present currents $\eta_{\mu\nu}(x)$ differ from that of Refs.\cite{Tetraquark-Zb-QCDSR,Chen-Zhu}.

 In the following,  we briefly outline  the operator product expansion for the correlation functions $\Pi_{\mu\nu\alpha\beta}(p)$ and $\Pi(p)$ in perturbative QCD.  We contract the $s$  and $Q$ quark fields in the correlation functions
$\Pi_{\mu\nu\alpha\beta}(p)$ and $\Pi(p)$ with Wick theorem, and obtain the results:
\begin{eqnarray}
\Pi_{\mu\nu\alpha\beta}(p)&=&\frac{i}{2}\int d^4x e^{ip \cdot x}   \left\{{\rm Tr}\left[ \gamma_{\mu}S_Q^{ij}(x)\gamma_{\alpha} S^{ji}(-x)\right] {\rm Tr}\left[ \gamma_{\nu} S^{mn}(x)\gamma_{\beta}  S_Q^{nm}(-x)\right] \right. \nonumber\\
&&+{\rm Tr}\left[ \gamma_{\nu}S_Q^{ij}(x)\gamma_{\beta} S^{ji}(-x)\right] {\rm Tr}\left[ \gamma_{\mu} S^{mn}(x)\gamma_{\alpha}  S_Q^{nm}(-x)\right] \nonumber\\
&&\pm{\rm Tr}\left[ \gamma_{\nu}S_Q^{ij}(x)\gamma_{\alpha} S^{ji}(-x)\right] {\rm Tr}\left[ \gamma_{\mu} S^{mn}(x)\gamma_{\beta}  S_Q^{nm}(-x)\right]\nonumber\\
 &&\left.\pm {\rm Tr}\left[ \gamma_{\mu}S_Q^{ij}(x)\gamma_{\beta} S^{ji}(-x)\right] {\rm Tr}\left[ \gamma_{\nu} S^{mn}(x)\gamma_{\alpha}  S_Q^{nm}(-x)\right] \right\} \, , \\
 \Pi(p)&=&i\int d^4x e^{ip \cdot x}  {\rm Tr}\left[ \gamma_{\mu}S_Q^{ij}(x)\gamma_{\alpha} S^{ji}(-x)\right] {\rm Tr}\left[ \gamma^{\mu} S^{mn}(x)\gamma^{\alpha}  S_Q^{nm}(-x)\right]   \, ,
\end{eqnarray}
where the $\pm$ correspond to $\pm$ charge conjugations  respectively,
 the $S^{ij}(x)$ and $S_Q^{ij}(x)$ are the full $s$ and $Q$ quark propagators respectively,
 \begin{eqnarray}
S^{ij}(x)&=& \frac{i\delta_{ij}\!\not\!{x}}{ 2\pi^2x^4}
-\frac{\delta_{ij}m_s}{4\pi^2x^2}-\frac{\delta_{ij}\langle
\bar{s}s\rangle}{12} +\frac{i\delta_{ij}\!\not\!{x}m_s
\langle\bar{s}s\rangle}{48}-\frac{\delta_{ij}x^2\langle \bar{s}g_s\sigma Gs\rangle}{192}+\frac{i\delta_{ij}x^2\!\not\!{x} m_s\langle \bar{s}g_s\sigma
 Gs\rangle }{1152}\nonumber\\
&& -\frac{ig_s G^{a}_{\alpha\beta}t^a_{ij}(\!\not\!{x}
\sigma^{\alpha\beta}+\sigma^{\alpha\beta} \!\not\!{x})}{32\pi^2x^2} -\frac{i\delta_{ij}x^2\!\not\!{x}g_s^2\langle \bar{s} s\rangle^2}{7776} -\frac{\delta_{ij}x^4\langle \bar{s}s \rangle\langle g_s^2 GG\rangle}{27648}-\frac{1}{8}\langle\bar{s}_j\sigma^{\mu\nu}s_i \rangle \sigma_{\mu\nu} \nonumber\\
&&   -\frac{1}{4}\langle\bar{s}_j\gamma^{\mu}s_i\rangle \gamma_{\mu }+\cdots \, ,
\end{eqnarray}
\begin{eqnarray}
S_Q^{ij}(x)&=&\frac{i}{(2\pi)^4}\int d^4k e^{-ik \cdot x} \left\{
\frac{\delta_{ij}}{\!\not\!{k}-m_Q}
-\frac{g_sG^n_{\alpha\beta}t^n_{ij}}{4}\frac{\sigma^{\alpha\beta}(\!\not\!{k}+m_Q)+(\!\not\!{k}+m_Q)
\sigma^{\alpha\beta}}{(k^2-m_Q^2)^2}\right.\nonumber\\
&&\left. +\frac{g_s D_\alpha G^n_{\beta\lambda}t^n_{ij}(f^{\lambda\beta\alpha}+f^{\lambda\alpha\beta}) }{3(k^2-m_Q^2)^4}-\frac{g_s^2 (t^at^b)_{ij} G^a_{\alpha\beta}G^b_{\mu\nu}(f^{\alpha\beta\mu\nu}+f^{\alpha\mu\beta\nu}+f^{\alpha\mu\nu\beta}) }{4(k^2-m_Q^2)^5}+\cdots\right\} \, ,\nonumber\\
f^{\lambda\alpha\beta}&=&(\!\not\!{k}+m_Q)\gamma^\lambda(\!\not\!{k}+m_Q)\gamma^\alpha(\!\not\!{k}+m_Q)\gamma^\beta(\!\not\!{k}+m_Q)\, ,\nonumber\\
f^{\alpha\beta\mu\nu}&=&(\!\not\!{k}+m_Q)\gamma^\alpha(\!\not\!{k}+m_Q)\gamma^\beta(\!\not\!{k}+m_Q)\gamma^\mu(\!\not\!{k}+m_Q)\gamma^\nu(\!\not\!{k}+m_Q)\, ,
\end{eqnarray}
and  $t^n=\frac{\lambda^n}{2}$, the $\lambda^n$ is the Gell-Mann matrix,  $D_\alpha=\partial_\alpha-ig_sG^n_\alpha t^n$ \cite{Reinders85}, then compute  the integrals both in the coordinate and momentum spaces,  and obtain the correlation functions $\Pi_{\mu\nu\alpha\beta}(p)$ and  $\Pi(p)$  therefore the QCD spectral densities \footnote{It is convenient  to introduce the external fields $\bar{\chi}$, $\chi$, $A_\alpha^a$ and additional Lagrangian $\Delta \mathcal{L}$
\begin{eqnarray}
\Delta {\mathcal{L}}&=&\bar{s}(x)\left(i\gamma^\mu \partial_\mu-m_s\right)\chi(x)+\bar{\chi}(x)\left(i\gamma^\mu \partial_\mu-m_s\right) s(x)+g_s \bar{s}(x)\gamma^\mu t^a s(x)A_\mu^a(x)+\cdots\, , \nonumber
\end{eqnarray}
in carrying out the operator product expansion \cite{Reinders85,external}. We expand the heavy and light quark propagators $S^Q_{ij}$ and $S_{ij}$ in terms of the
external fields $\bar{\chi}$, $\chi$ and $A_\alpha^a$,
\begin{eqnarray}
S^Q_{ij}\left(x,\bar{\chi},\chi,A_\alpha^a\right)&=&\frac{i}{(2\pi)^4}\int d^4k e^{-ik \cdot x} \left\{
\frac{\delta_{ij}}{\!\not\!{k}-m_Q}
-\frac{g_s A^a_{\alpha\beta}t^a_{ij}}{4}\frac{\sigma^{\alpha\beta}(\!\not\!{k}+m_Q)+(\!\not\!{k}+m_Q)
\sigma^{\alpha\beta}}{(k^2-m_Q^2)^2}+\cdots\right\} \, , \nonumber
\end{eqnarray}
\begin{eqnarray}
S_{ij}\left(x,\bar{\chi},\chi,A_\mu^a\right)&=& \frac{i\delta_{ij}\!\not\!{x}}{ 2\pi^2x^4}+\chi^i(x)\bar{\chi}^j(0) -\frac{ig_s A^{a}_{\alpha\beta}t^a_{ij}\left(\!\not\!{x}
\sigma^{\alpha\beta}+\sigma^{\alpha\beta} \!\not\!{x}\right)}{32\pi^2x^2}+\cdots \, ,\nonumber
\end{eqnarray}
where $A^a_{\alpha\beta}=\partial_\alpha A^a_\beta-\partial_\beta A^a_\alpha+g_s f^{abc}A_\alpha^b A_\beta^c$.
Then the  correlation functions $\Pi(p)$ can be written as
\begin{eqnarray}
\Pi(p)&=& \sum_{n=0}^\infty {\mathcal{C}}_n(p,\mu)\,\, {\mathcal{O}}_n\left(\bar{\chi},\chi,A_\alpha^a,\mu\right)\, , \nonumber
\end{eqnarray}
in the external fields $\bar{\chi}$, $\chi$ and $A_\alpha^a$, where the ${\mathcal{C}}_n(p,\mu)$ are the Wilson's coefficients, the ${\mathcal{O}}_n\left(\bar{\chi},\chi,A_\alpha^a,\mu\right)$ are operators   characterized by their dimensions $n$.
We choose the energy scale $\mu\gg \Lambda_{QCD}$, the Wilson coefficients ${\mathcal{C}}_n(p^2,\mu)$ depend only on short-distance dynamics,
  and the perturbative calculations make sense.
If we neglect the perturbative (or radiative) corrections, the
 operators  ${\mathcal{O}}_n\left(\bar{\chi},\chi,A_\alpha^a,\mu\right)$ can also be counted by the orders of the fine  constant $\alpha_s(\mu)=\frac{g_s^2(\mu)}{4\pi}$, ${\mathcal{O}}\left(\alpha_s^k\right)$,
  with $k=0, \frac{1}{2}, 1, \frac{3}{2}$, etc. In this article, we take the truncations $n\leq 10$ and $k\leq 1$, and
  factorize the higher dimensional operators into non-factorizable
   low dimensional  operators with the same quantum numbers of the vacuum.
  Taking the following replacements
  \begin{eqnarray}
  {\mathcal{O}}_n\left(\bar{\chi},\chi,A_\alpha^a,\mu\right) &\to& \langle{\mathcal{O}}_n\left(\bar{s},s,G_\alpha^a,\mu\right)\rangle \, ,\nonumber
  \end{eqnarray}
  we obtain the correlation functions at the level of quark-gluon degrees of freedom. For example,
  \begin{eqnarray}
    \chi^i(x)\bar{\chi}^j(0)  &=& -\frac{\delta_{ij}
\bar{\chi}(0)\chi(0) }{12} -\frac{\delta_{ij}x^2  \bar{\chi}(0)g_s\sigma A(0)\chi(0) }{192}+\cdots\to -\frac{\delta_{ij}\langle
\bar{s}s\rangle}{12} -\frac{\delta_{ij}x^2\langle \bar{s}g_s\sigma Gs\rangle}{192}+\cdots\, . \nonumber
  \end{eqnarray}
   For simplicity, we often  take the following replacements,
  \begin{eqnarray}
  S^Q_{ij}\left(x,\bar{\chi},\chi,A_\alpha^a\right)&\to&S^Q_{ij}\left(x,\bar{s},s,G_\alpha^a\right)\, , \nonumber\\
  S_{ij}\left(x,\bar{\chi},\chi,A_\alpha^a\right)&\to&S_{ij}\left(x,\bar{s},s,G_\alpha^a\right)\, , \nonumber\\
 {\mathcal{O}}_n\left(\bar{\chi},\chi,A_\alpha^a\right) &\to& \langle {\mathcal{O}}_n\left(\bar{s},s,G_\alpha^a\right)\rangle\, , \nonumber
  \end{eqnarray}
directly in calculations by neglecting some intermediate steps, and resort to the routine  taken in this article.}.
In Eq.(18), we retain the terms $\langle\bar{s}_j\sigma_{\mu\nu}s_i \rangle$ and $\langle\bar{s}_j\gamma_{\mu}s_i\rangle$ originate from the Fierz re-ordering of the $\langle s_i \bar{s}_j\rangle$ to  absorb the gluons  emitted from the heavy quark lines to form $\langle\bar{s}_j g_s G^a_{\alpha\beta} t^a_{mn}\sigma_{\mu\nu} s_i \rangle$ and $\langle\bar{s}_j\gamma_{\mu}s_ig_s D_\nu G^a_{\alpha\beta}t^a_{mn}\rangle$ so as to extract the mixed condensate and four-quark condensates $\langle\bar{s}g_s\sigma G s\rangle$ and $g_s^2\langle\bar{s}s\rangle^2$, respectively.
The $s$-quark fields $s(x)$, $\bar{s}(x)$ and gluon field $G^a_\mu(x)$  can be expanded in terms of the Taylor series of  covariant derivatives,
\begin{eqnarray}
s(x)&=&\sum_{n=0}^{\infty} \frac{1}{n!}\,x^{\mu_1}x^{\mu_2}\cdots x^{\mu_n}\, D_{\mu_1}D_{\mu_2} \cdots D_{\mu_n}\, s(0)\, , \nonumber\\
\bar{s}(x)&=&\sum_{n=0}^{\infty} \frac{1}{n!}\,x^{\mu_1}x^{\mu_2}\cdots x^{\mu_n}\, \bar{s}(0)\, D^{\dagger}_{\mu_1}D^{\dagger}_{\mu_2} \cdots D^{\dagger}_{\mu_n} \, , \nonumber\\
G^a_\mu(x)&=&\sum_{n=0}^{\infty} \frac{1}{n!(n+2)}\,x^{\rho}x^{\mu_1}x^{\mu_2}\cdots x^{\mu_n}\, D_{\mu_1}D_{\mu_2} \cdots D_{\mu_n}\, G^a_{\rho\mu}(0)\, .
\end{eqnarray}
The bilinear fields $s_\alpha(x) \bar{s}_\beta(0)$ can be re-arranged into the following form in the Dirac spinor space,
\begin{eqnarray}
s_\alpha(x) \bar{s}_\beta(0)&=&-\frac{1}{4}
\delta_{\alpha\beta}\bar{s}(0)s(x)
-\frac{1}{4}(\gamma^\mu)_{\alpha\beta}\bar{s}(0)\gamma_\mu
s(x) -\frac{1}{8}(\sigma^{\mu\nu})_{\alpha\beta}\bar{s}(0)\sigma_{\mu\nu}s(x) \nonumber\\
&&+\frac{1}{4}(\gamma^\mu
\gamma_5)_{\alpha\beta}\bar{s}(0)\gamma_\mu \gamma_5 s(x)+\frac{1}{4}(i \gamma_5)_{\alpha\beta}\bar{s}(0)i\gamma_5 s(x) \, .
\end{eqnarray}
The vacuum condensates $\langle \bar{s}g_s\sigma Gs\rangle$, $g_s^2\langle \bar{s} s\rangle^2$ and $\langle \bar{s}s \rangle\langle g_s^2 GG\rangle$ in the full $s$-quark propagator originate  from the vacuum expectations of the operators
$\bar{s}(0)\sigma^{\mu\nu} D^{\alpha}D^{\beta}s(0)$, $\bar{s}(0)\gamma^{\mu} D^{\alpha}D^{\beta}D^{\lambda}s(0)$ and $\bar{s}(0) D^{\alpha}D^{\beta}D^{\lambda}D^{\tau}s(0)$, respectively.
We take into account the formulas $\left[D_\alpha, D_\beta \right]=-ig_sG_{\alpha\beta}$ and $D^\alpha G^a_{\alpha\mu}=-g_s \left(\bar{u}\gamma_{\mu}t^a u+\bar{d}\gamma_{\mu}t^a d +\bar{s}\gamma_{\mu}t^a s\right)$, then the terms with $n>4$ in the Taylor expansion of the $s(x)$ and $\bar{s}(x)$ are of the order ${\mathcal{O}}(\alpha_s^k)$ with $k>1$, and have no contribution in the present truncation. The operators $g_sG^n_{\alpha\beta}$, $g_s D_\alpha G^n_{\beta\lambda}$ and
$g_s^2  G^a_{\alpha\beta}G^b_{\mu\nu}$ in the full $Q$-quark propagator are of the order ${\mathcal{O}}(\alpha_s^k)$ with $k=\frac{1}{2}, \, 1$ and $1$, respectively.
The terms with $n>1$ in the Taylor expansion of the $G^a_\mu(x)$ are of the order ${\mathcal{O}}(\alpha_s^k)$ with $k>1$, and have no contribution in the present truncation. In this article, we take  the truncation ${\mathcal{O}}(\alpha_s^k)$ with $k\leq 1$, the operators therefore the vacuum condensates have the dimensions less than or equal 10.

 Once the analytical expressions  are obtained,  we can take the
quark-hadron duality below the continuum thresholds  $s_0$ and perform Borel transform  with respect to
the variable $P^2=-p^2$ to obtain  the following QCD sum rules:
\begin{eqnarray}
\lambda^2_{Y/Z}\, \exp\left(-\frac{M^2_{Y/Z}}{T^2}\right)= \int_{4m_Q^2}^{s_0} ds\, \rho(s) \, \exp\left(-\frac{s}{T^2}\right) \, ,
\end{eqnarray}
where
\begin{eqnarray}
\rho(s)&=&\rho_{0}(s)+\rho_{3}(s) +\rho_{4}(s)+\rho_{5}(s)+\rho_{6}(s)+\rho_{7}(s) +\rho_{8}(s)+\rho_{10}(s)\, ,
\end{eqnarray}

\begin{eqnarray}
\rho^{J=2}_{0}(s)&=&\frac{1}{20480\pi^6}\int_{y_i}^{y_f}dy \int_{z_i}^{1-y}dz \, yz\, (1-y-z)^3\left(s-\overline{m}_Q^2\right)^2\left(293s^2-190s\overline{m}_Q^2+17\overline{m}_Q^4 \right)  \nonumber\\
&&+\frac{3}{20480\pi^6} \int_{y_i}^{y_f}dy \int_{z_i}^{1-y}dz \, yz \,(1-y-z)^2\left(s-\overline{m}_Q^2\right)^4    \nonumber\\
&&+\frac{3m_sm_Q}{512\pi^6}\int_{y_i}^{y_f}dy \int_{z_i}^{1-y}dz \, (y+z)\, (1-y-z)^2\left(s-\overline{m}_Q^2\right)^2\left(4s-\overline{m}_Q^2 \right) \, ,
\end{eqnarray}

\begin{eqnarray}
\rho_{3}^{J=2}(s)&=&-\frac{3m_Q\langle \bar{s}s\rangle}{64\pi^4}\int_{y_i}^{y_f}dy \int_{z_i}^{1-y}dz \, (y+z)(1-y-z)\left(s-\overline{m}_Q^2\right)\left(3s-\overline{m}_Q^2\right)  \nonumber\\
&&+\frac{3m_s\langle \bar{s}s\rangle}{640\pi^4}\int_{y_i}^{y_f}dy \int_{z_i}^{1-y}dz \, yz\, (1-y-z)\left(115s^2-112s\overline{m}_Q^2+17\overline{m}_Q^4 \right) \nonumber\\
&&+\frac{3m_s\langle \bar{s}s\rangle}{640\pi^4}\int_{y_i}^{y_f}dy \int_{z_i}^{1-y}dz \, yz \left(s-\overline{m}_Q^2 \right)^2 \nonumber\\
&&-\frac{3m_sm_Q^2\langle \bar{s}s\rangle}{16\pi^4}\int_{y_i}^{y_f}dy \int_{z_i}^{1-y}dz  \left(s-\overline{m}_Q^2 \right) \, ,
\end{eqnarray}

\begin{eqnarray}
\rho_{4}^{J=2}(s)&=&-\frac{m_Q^2}{15360\pi^4} \langle\frac{\alpha_s GG}{\pi}\rangle\int_{y_i}^{y_f}dy \int_{z_i}^{1-y}dz \left( \frac{z}{y^2}+\frac{y}{z^2}\right)(1-y-z)^3 \nonumber\\
&&\left\{ 56s-17\overline{m}_Q^2+10s^2\delta\left(s-\overline{m}_Q^2\right)\right\} \nonumber\\
&&-\frac{m_Q^2}{5120\pi^4}\langle\frac{\alpha_s GG}{\pi}\rangle\int_{y_i}^{y_f}dy \int_{z_i}^{1-y}dz \left(\frac{z}{y^2}+\frac{y}{z^2} \right) (1-y-z)^2 \left(s-\overline{m}_Q^2\right) \nonumber\\
&&-\frac{1}{10240\pi^4} \langle\frac{\alpha_s GG}{\pi}\rangle\int_{y_i}^{y_f}dy \int_{z_i}^{1-y}dz \left( y+z\right)(1-y-z)^2 \left( 185s^2-208s\overline{m}_Q^2+43\overline{m}_Q^4\right) \nonumber\\
&&+\frac{1}{5120\pi^4} \langle\frac{\alpha_s GG}{\pi}\rangle\int_{y_i}^{y_f}dy \int_{z_i}^{1-y}dz \left( y+z\right)(1-y-z) \left( s-\overline{m}_Q^2\right)^2
 \, ,
\end{eqnarray}

\begin{eqnarray}
\rho^{J=2}_{5}(s)&=&\frac{3m_Q\langle \bar{s}g_s\sigma Gs\rangle}{128\pi^4}\int_{y_i}^{y_f}dy \int_{z_i}^{1-y}dz  \, (y+z) \left(2s-\overline{m}_Q^2 \right) \nonumber\\
&&-\frac{m_s\langle \bar{s}g_s\sigma Gs\rangle}{640\pi^4}\int_{y_i}^{y_f}dy \int_{z_i}^{1-y}dz  \,  yz \left\{56s-17\overline{m}_Q^2+10s^2 \delta\left(s-\overline{m}_Q^2 \right)\right\}    \nonumber\\
&&-\frac{m_s\langle \bar{s}g_s\sigma Gs\rangle}{640\pi^4}\int_{y_i}^{y_f}dy    \,  y(1-y) \left(s-\widetilde{m}_Q^2 \right) +\frac{3m_s m_Q^2\langle \bar{s}g_s\sigma Gs\rangle}{64\pi^4}\int_{y_i}^{y_f}dy  \, ,
\end{eqnarray}

\begin{eqnarray}
\rho_{6}^{J=2}(s)&=&\frac{m_Q^2\langle\bar{s}s\rangle^2}{8\pi^2}\int_{y_i}^{y_f}dy -\frac{m_s m_Q\langle\bar{s}s\rangle^2}{16\pi^2}\int_{y_i}^{y_f}dy \left\{1+s \delta\left(s-\widetilde{m}_Q^2 \right)\right\} \nonumber\\
 &&+\frac{g_s^2\langle\bar{s}s\rangle^2}{4320\pi^4}\int_{y_i}^{y_f}dy \int_{z_i}^{1-y}dz\, yz \left\{56s-17\overline{m}_Q^2 +10\overline{m}_Q^4\delta\left(s-\overline{m}_Q^2 \right)\right\}\nonumber\\
&&+\frac{g_s^2\langle\bar{s}s\rangle^2}{4320\pi^4}\int_{y_i}^{y_f}dy \,y(1-y)\left(s-\widetilde{m}_Q^2 \right)  \nonumber\\
&&-\frac{g_s^2\langle\bar{s}s\rangle^2}{6480\pi^4}\int_{y_i}^{y_f}dy \int_{z_i}^{1-y}dz \, (1-y-z)\left\{ 45\left(\frac{z}{y}+\frac{y}{z} \right)\left(2s-\overline{m}_Q^2 \right)+\left(\frac{z}{y^2}+\frac{y}{z^2} \right)\right.\nonumber\\
&&\left.m_Q^2\left[ 19+20\overline{m}_Q^2\delta\left(s-\overline{m}_Q^2 \right)\right]+(y+z)\left[18\left(3s-\overline{m}_Q^2\right)+10\overline{m}_Q^4\delta\left(s-\overline{m}_Q^2\right) \right] \right\}\, ,
\end{eqnarray}

\begin{eqnarray}
\rho_7^{J=2}(s)&=&\frac{m_Q^3\langle\bar{s}s\rangle}{192\pi^2 T^2}\langle\frac{\alpha_sGG}{\pi}\rangle\int_{y_i}^{y_f}dy \int_{z_i}^{1-y}dz \left(\frac{y}{z^3}+\frac{z}{y^3}+\frac{1}{y^2}+\frac{1}{z^2}\right)(1-y-z)\, \overline{m}_Q^2 \, \delta\left(s-\overline{m}_Q^2\right)\nonumber\\
&&-\frac{m_Q\langle\bar{s}s\rangle}{64\pi^2}\langle\frac{\alpha_sGG}{\pi}\rangle\int_{y_i}^{y_f}dy \int_{z_i}^{1-y}dz \left(\frac{y}{z^2}+\frac{z}{y^2}\right)(1-y-z)  \left\{1+\overline{m}_Q^2\delta\left(s-\overline{m}_Q^2\right) \right\}\nonumber\\
&&+\frac{m_Q\langle\bar{s}s\rangle}{32\pi^2}\langle\frac{\alpha_sGG}{\pi}\rangle\int_{y_i}^{y_f}dy \int_{z_i}^{1-y}dz\left\{1+\frac{\overline{m}_Q^2}{3}\delta\left(s-\overline{m}_Q^2\right) \right\} \nonumber\\
&&-\frac{m_Q\langle\bar{s}s\rangle}{384\pi^2}\langle\frac{\alpha_sGG}{\pi}\rangle\int_{y_i}^{y_f}dy \left\{1+ \widetilde{m}_Q^2 \, \delta \left(s-\widetilde{m}_Q^2\right) \right\}\, ,
\end{eqnarray}

\begin{eqnarray}
\rho_8^{J=2}(s)&=&-\frac{m_Q^2\langle\bar{s}s\rangle\langle\bar{s}g_s\sigma Gs\rangle}{16\pi^2}\int_0^1 dy \left(1+\frac{\widetilde{m}_Q^2}{T^2} \right)\delta\left(s-\widetilde{m}_Q^2\right) \, ,
\end{eqnarray}

\begin{eqnarray}
\rho_{10}^{J=2}(s)&=&\frac{m_Q^2\langle\bar{s}g_s\sigma Gs\rangle^2}{128\pi^2T^6}\int_0^1 dy \, \widetilde{m}_Q^4 \, \delta \left( s-\widetilde{m}_Q^2\right)
\nonumber \\
&&-\frac{m_Q^4\langle\bar{s}s\rangle^2}{144T^4}\langle\frac{\alpha_sGG}{\pi}\rangle\int_0^1 dy  \left\{ \frac{1}{y^3}+\frac{1}{(1-y)^3}\right\} \delta\left( s-\widetilde{m}_Q^2\right)\nonumber\\
&&+\frac{m_Q^2\langle\bar{s}s\rangle^2}{48T^2}\langle\frac{\alpha_sGG}{\pi}\rangle\int_0^1 dy  \left\{ \frac{1}{y^2}+\frac{1}{(1-y)^2}\right\} \delta\left( s-\widetilde{m}_Q^2\right)\nonumber\\
&&+\frac{\langle\bar{s}g_s\sigma Gs\rangle^2}{5184 \pi^2T^2} \int_0^1 dy   \, \widetilde{m}_Q^2 \delta\left( s-\widetilde{m}_Q^2\right)\nonumber\\
&&+\frac{m_Q^2\langle\bar{s} s\rangle^2}{144 T^6}\langle\frac{\alpha_sGG}{\pi}\rangle\int_0^1 dy \, \widetilde{m}_Q^4 \, \delta \left( s-\widetilde{m}_Q^2\right) \, ,
\end{eqnarray}

\begin{eqnarray}
\rho^{J=1}_{0}(s)&=&\frac{1}{4096\pi^6s}\int_{y_i}^{y_f}dy \int_{z_i}^{1-y}dz \, yz\, (1-y-z)^3\left(s-\overline{m}_Q^2\right)^2\left(49s^2-30s\overline{m}_Q^2+\overline{m}_Q^4 \right)  \nonumber\\
&&+\frac{1}{4096\pi^6s} \int_{y_i}^{y_f}dy \int_{z_i}^{1-y}dz \, yz \,(1-y-z)^2\left(s-\overline{m}_Q^2\right)^3\left(3s+\overline{m}_Q^2\right)   \nonumber\\
&&+\frac{9m_sm_Q}{1024\pi^6} \int_{y_i}^{y_f}dy \int_{z_i}^{1-y}dz \, (y+z) \,(1-y-z)^2\left(s-\overline{m}_Q^2\right)^2 \, ,
\end{eqnarray}

\begin{eqnarray}
\rho_{3}^{J=1}(s)&=&-\frac{3 m_Q\langle \bar{s}s\rangle}{64\pi^4}\int_{y_i}^{y_f}dy \int_{z_i}^{1-y}dz \, (y+z)(1-y-z)\left(s-\overline{m}_Q^2\right)  \nonumber\\
&&+\frac{m_s\langle \bar{s}s\rangle}{128\pi^4s}\int_{y_i}^{y_f}dy \int_{z_i}^{1-y}dz \, yz\, (1-y-z)\left(55s^2-48s\overline{m}_Q^2+3\overline{m}_Q^4 \right) \nonumber\\
&&+\frac{m_s\langle \bar{s}s\rangle}{128\pi^4s}\int_{y_i}^{y_f}dy \int_{z_i}^{1-y}dz \, yz\, \left(s-\overline{m}_Q^2 \right) \left(s+\overline{m}_Q^2 \right) \, ,
\end{eqnarray}

\begin{eqnarray}
\rho_{4}^{J=1}(s)&=&-\frac{m_Q^2}{3072\pi^4s} \langle\frac{\alpha_s GG}{\pi}\rangle\int_{y_i}^{y_f}dy \int_{z_i}^{1-y}dz \left( \frac{z}{y^2}+\frac{y}{z^2}\right)(1-y-z)^3 \nonumber\\
&&\left\{ 8s-\overline{m}_Q^2+\frac{5\overline{m}_Q^4}{3}\delta\left(s-\overline{m}_Q^2\right)\right\} \nonumber\\
&&-\frac{m_Q^2}{3072\pi^4s}\langle\frac{\alpha_s GG}{\pi}\rangle\int_{y_i}^{y_f}dy \int_{z_i}^{1-y}dz \left(\frac{z}{y^2}+\frac{y}{z^2} \right) (1-y-z)^2 \, \overline{m}_Q^2 \nonumber\\
&&-\frac{1}{6144\pi^4s} \langle\frac{\alpha_s GG}{\pi}\rangle\int_{y_i}^{y_f}dy \int_{z_i}^{1-y}dz \left( y+z\right)(1-y-z)^2 \left( 5s^2-3\overline{m}_Q^4\right) \nonumber\\
&&+\frac{1}{3072\pi^4s} \langle\frac{\alpha_s GG}{\pi}\rangle\int_{y_i}^{y_f}dy \int_{z_i}^{1-y}dz \left( y+z\right)(1-y-z) \left( s^2-\overline{m}_Q^4\right) \, ,
\end{eqnarray}

\begin{eqnarray}
\rho^{J=1}_{5}(s)&=&\frac{3m_Q\langle \bar{s}g_s\sigma Gs\rangle}{256\pi^4}\int_{y_i}^{y_f}dy \int_{z_i}^{1-y}dz  \, (y+z)\nonumber\\
&&-\frac{m_s\langle \bar{s}g_s\sigma Gs\rangle}{128\pi^4s}\int_{y_i}^{y_f}dy \int_{z_i}^{1-y}dz  \,  yz \left\{8s-\overline{m}_Q^2+\frac{5s^2}{3} \delta\left(s-\overline{m}_Q^2 \right)\right\}    \nonumber\\
&&-\frac{m_s\langle \bar{s}g_s\sigma Gs\rangle}{384\pi^4s}\int_{y_i}^{y_f}dy \, y(1-y) \, \widetilde{m}_Q^2          \, ,
\end{eqnarray}

\begin{eqnarray}
\rho_{6}^{J=1}(s)&=&-\frac{m_sm_Q\langle \bar{s}s\rangle^2}{32\pi^2}\int_0^1 dy \, \delta \left(s-\widetilde{m}_Q^2 \right) \, , \nonumber\\
&&+\frac{g_s^2\langle\bar{s}s\rangle^2}{864\pi^4s}\int_{y_i}^{y_f}dy \int_{z_i}^{1-y}dz\, yz \left\{8s-\overline{m}_Q^2 +\frac{5\overline{m}_Q^4}{3}\delta\left(s-\overline{m}_Q^2 \right)\right\}\nonumber\\
&&+\frac{g_s^2\langle\bar{s}s\rangle^2}{2592\pi^4s}\int_{y_i}^{y_f}dy \,y(1-y) \,\widetilde{m}_Q^2   \nonumber\\
&&-\frac{g_s^2\langle\bar{s}s\rangle^2}{864\pi^4}\int_{y_i}^{y_f}dy \int_{z_i}^{1-y}dz \, (1-y-z)\left\{ 3\left(\frac{z}{y}+\frac{y}{z} \right)  \right.\nonumber\\
&&\left. +\left(\frac{z}{y^2}+\frac{y}{z^2} \right)m_Q^2  \delta\left(s-\overline{m}_Q^2 \right)+(y+z)\left[8+2\overline{m}_Q^2\delta\left(s-\overline{m}_Q^2\right) \right] \right\} \, ,
\end{eqnarray}

\begin{eqnarray}
\rho_7^{J=1}(s)&=&\frac{m_Q^3\langle\bar{s}s\rangle}{384\pi^2s}\langle\frac{\alpha_sGG}{\pi}\rangle\int_{y_i}^{y_f}dy \int_{z_i}^{1-y}dz \left(\frac{y}{z^3}+\frac{z}{y^3}+\frac{1}{y^2}+\frac{1}{z^2}\right)(1-y-z)\left( 1+\frac{s}{T^2}\right) \nonumber\\
&&\delta\left(s-\overline{m}_Q^2\right)-\frac{m_Q\langle\bar{s}s\rangle}{128\pi^2}\langle\frac{\alpha_sGG}{\pi}\rangle\int_{y_i}^{y_f}dy \int_{z_i}^{1-y}dz \left(\frac{y}{z^2}+\frac{z}{y^2}\right)(1-y-z)  \delta\left(s-\overline{m}_Q^2\right)\nonumber\\
&&-\frac{m_Q\langle\bar{s}s\rangle}{192\pi^2}\langle\frac{\alpha_sGG}{\pi}\rangle\int_{y_i}^{y_f}dy \int_{z_i}^{1-y}dz \, \delta\left(s-\overline{m}_Q^2\right) \nonumber\\
&&-\frac{m_Q\langle\bar{s}s\rangle}{768\pi^2}\langle\frac{\alpha_sGG}{\pi}\rangle\int_{y_i}^{y_f}dy \int_{z_i}^{1-y}dz \,\delta\left(s-\overline{m}_Q^2\right)  \, ,
\end{eqnarray}

\begin{eqnarray}
\rho^{J=0}_{0}(s)&=&\frac{3}{1024\pi^6}\int_{y_i}^{y_f}dy \int_{z_i}^{1-y}dz \, yz\, (1-y-z)^3\left(s-\overline{m}_Q^2\right)^2\left(7s^2-6s\overline{m}_Q^2+\overline{m}_Q^4 \right)  \nonumber\\
&&+\frac{3}{1024\pi^6} \int_{y_i}^{y_f}dy \int_{z_i}^{1-y}dz \, yz \,(1-y-z)^2\left(s-\overline{m}_Q^2\right)^3 \left(3s-\overline{m}_Q^2\right) \nonumber\\
&&+\frac{3m_sm_Q}{512\pi^6} \int_{y_i}^{y_f}dy \int_{z_i}^{1-y}dz \, (y+z) \,(1-y-z)^2\left(s-\overline{m}_Q^2\right)^2 \left(5s-2\overline{m}_Q^2\right)  \, ,
\end{eqnarray}

\begin{eqnarray}
\rho_{3}^{J=0}(s)&=&-\frac{3m_Q\langle \bar{s}s\rangle}{32\pi^4}\int_{y_i}^{y_f}dy \int_{z_i}^{1-y}dz \, (y+z)(1-y-z)\left(s-\overline{m}_Q^2\right)\left(2s-\overline{m}_Q^2\right)  \nonumber\\
&&+\frac{3m_s\langle \bar{s}s\rangle}{32\pi^4}\int_{y_i}^{y_f}dy \int_{z_i}^{1-y}dz \, yz\,(1-y-z)\left(10s^2-12s\overline{m}_Q^2+3\overline{m}_Q^4 \right)\nonumber\\
&&+\frac{3m_s\langle \bar{s}s\rangle}{32\pi^4}\int_{y_i}^{y_f}dy \int_{z_i}^{1-y}dz \, yz\, \left(s-\overline{m}_Q^2\right)\left(2s-\overline{m}_Q^2\right)\nonumber\\
&&-\frac{3m_sm_Q^2\langle \bar{s}s\rangle}{8\pi^4}\int_{y_i}^{y_f}dy \, \left(s-\widetilde{m}_Q^2\right)   \, ,
\end{eqnarray}

\begin{eqnarray}
\rho_{4}^{J=0}(s)&=&-\frac{m_Q^2}{256\pi^4} \langle\frac{\alpha_s GG}{\pi}\rangle\int_{y_i}^{y_f}dy \int_{z_i}^{1-y}dz \left( \frac{z}{y^2}+\frac{y}{z^2}\right)(1-y-z)^3 \nonumber\\
&&\left\{ 2s-\overline{m}_Q^2+\frac{\overline{m}_Q^4}{6}\delta\left(s-\overline{m}_Q^2\right)\right\} \nonumber\\
&&-\frac{m_Q^2}{512\pi^4}\langle\frac{\alpha_s GG}{\pi}\rangle\int_{y_i}^{y_f}dy \int_{z_i}^{1-y}dz \left(\frac{z}{y^2}+\frac{y}{z^2} \right) (1-y-z)^2 \left(3s-2\overline{m}_Q^2\right) \nonumber\\
&&-\frac{1}{512\pi^4} \langle\frac{\alpha_s GG}{\pi}\rangle\int_{y_i}^{y_f}dy \int_{z_i}^{1-y}dz \left( y+z\right)(1-y-z)^2 \left( 10s^2-12s\overline{m}_Q^2+3\overline{m}_Q^4\right) \nonumber\\
&&+\frac{1}{256\pi^4} \langle\frac{\alpha_s GG}{\pi}\rangle\int_{y_i}^{y_f}dy \int_{z_i}^{1-y}dz \left( y+z\right)(1-y-z) \left( s-\overline{m}_Q^2\right)\left( 2s-\overline{m}_Q^2\right) \, ,
\end{eqnarray}

\begin{eqnarray}
\rho^{J=0}_{5}(s)&=&\frac{3m_Q\langle \bar{s}g_s\sigma Gs\rangle}{128\pi^4}\int_{y_i}^{y_f}dy \int_{z_i}^{1-y}dz  \, (y+z) \left(3s-2\overline{m}_Q^2 \right) \nonumber\\
&&-\frac{3m_s\langle \bar{s}g_s\sigma Gs\rangle}{32\pi^4}\int_{y_i}^{y_f}dy \int_{z_i}^{1-y}dz  \,  yz\, \left\{2s-\overline{m}_Q^2+\frac{s}{6}\,\delta \left(s-\overline{m}_Q^2 \right) \right\}  \nonumber\\
&&-\frac{m_s\langle \bar{s}g_s\sigma Gs\rangle}{64\pi^4}\int_{y_i}^{y_f}dy   \,  y(1-y)\, \left(3s-2\overline{m}_Q^2 \right) \nonumber\\
&&+\frac{3m_sm_Q^2\langle \bar{s}g_s\sigma Gs\rangle}{32\pi^4}\int_{y_i}^{y_f}dy      \, ,
\end{eqnarray}

\begin{eqnarray}
\rho_{6}^{J=0}(s)&=&\frac{m_Q^2\langle\bar{s}s\rangle^2}{4\pi^2}\int_{y_i}^{y_f}dy  -\frac{m_sm_Q\langle\bar{s}s\rangle^2}{8\pi^2}\int_{y_i}^{y_f}dy \left\{1 +\frac{s}{2}\delta\left(s-\widetilde{m}_Q^2 \right)\right\}\nonumber\\
 &&+\frac{g_s^2\langle\bar{s}s\rangle^2}{72\pi^4}\int_{y_i}^{y_f}dy \int_{z_i}^{1-y}dz\, yz \left\{2s-\overline{m}_Q^2 +\frac{\overline{m}_Q^4}{6}\delta\left(s-\overline{m}_Q^2 \right)\right\}\nonumber\\
&&+\frac{g_s^2\langle\bar{s}s\rangle^2}{432\pi^4}\int_{y_i}^{y_f}dy \,y(1-y)\left(3s-2\widetilde{m}_Q^2 \right)  \nonumber\\
&&-\frac{g_s^2\langle\bar{s}s\rangle^2}{432\pi^4}\int_{y_i}^{y_f}dy \int_{z_i}^{1-y}dz \, (1-y-z)\left\{ 3\left(\frac{z}{y}+\frac{y}{z} \right)\left(3s-2\overline{m}_Q^2 \right)+\left(\frac{z}{y^2}+\frac{y}{z^2} \right)\right.\nonumber\\
&&\left.m_Q^2\left[ 2+ \overline{m}_Q^2\delta\left(s-\overline{m}_Q^2 \right)\right]+(y+z)\left[12\left(2s-\overline{m}_Q^2\right)+2\overline{m}_Q^4\delta\left(s-\overline{m}_Q^2\right) \right] \right\} \, ,
\end{eqnarray}

\begin{eqnarray}
\rho_7^{J=0}(s)&=&\frac{m_Q^3\langle\bar{s}s\rangle}{192\pi^2  }\langle\frac{\alpha_sGG}{\pi}\rangle\int_{y_i}^{y_f}dy \int_{z_i}^{1-y}dz \left(\frac{y}{z^3}+\frac{z}{y^3}+\frac{1}{y^2}+\frac{1}{z^2}\right)(1-y-z)\nonumber\\
&&\left(1+\frac{ s}{T^2}\right) \delta\left(s-\overline{m}_Q^2\right)\nonumber\\
&&-\frac{m_Q\langle\bar{s}s\rangle}{64\pi^2}\langle\frac{\alpha_sGG}{\pi}\rangle\int_{y_i}^{y_f}dy \int_{z_i}^{1-y}dz \left(\frac{y}{z^2}+\frac{z}{y^2}\right)(1-y-z)  \left\{2+\overline{m}_Q^2\delta\left(s-\overline{m}_Q^2\right) \right\}\nonumber\\
&&+\frac{m_Q\langle\bar{s}s\rangle}{32\pi^2}\langle\frac{\alpha_sGG}{\pi}\rangle\int_{y_i}^{y_f}dy \int_{z_i}^{1-y}dz\left\{2+ \overline{m}_Q^2 \delta\left(s-\overline{m}_Q^2\right) \right\} \nonumber\\
&&-\frac{m_Q\langle\bar{s}s\rangle}{384\pi^2}\langle\frac{\alpha_sGG}{\pi}\rangle\int_{y_i}^{y_f}dy \left\{2+ \widetilde{m}_Q^2 \, \delta \left(s-\widetilde{m}_Q^2\right) \right\}\, ,
\end{eqnarray}

\begin{eqnarray}
\rho_8^{J=0}(s)&=&-\frac{m_Q^2\langle\bar{s}s\rangle\langle\bar{s}g_s\sigma Gs\rangle}{8\pi^2}\int_0^1 dy \left(1+\frac{\widetilde{m}_Q^2}{T^2} \right)\delta\left(s-\widetilde{m}_Q^2\right) \, ,
\end{eqnarray}

\begin{eqnarray}
\rho_{10}^{J=0}(s)&=&\frac{m_Q^2\langle\bar{s}g_s\sigma Gs\rangle^2}{64\pi^2T^6}\int_0^1 dy \, \widetilde{m}_Q^4 \, \delta \left( s-\widetilde{m}_Q^2\right)
\nonumber \\
&&-\frac{m_Q^4\langle\bar{s}s\rangle^2}{72T^4}\langle\frac{\alpha_sGG}{\pi}\rangle\int_0^1 dy  \left\{ \frac{1}{y^3}+\frac{1}{(1-y)^3}\right\} \delta\left( s-\widetilde{m}_Q^2\right)\nonumber\\
&&+\frac{m_Q^2\langle\bar{s}s\rangle^2}{24T^2}\langle\frac{\alpha_sGG}{\pi}\rangle\int_0^1 dy  \left\{ \frac{1}{y^2}+\frac{1}{(1-y)^2}\right\} \delta\left( s-\widetilde{m}_Q^2\right)\nonumber\\
&&+\frac{m_Q^2\langle\bar{s}g_s\sigma Gs\rangle^2}{288 \pi^2T^2} \int_0^1 dy   \frac{1}{y(1-y)}   \delta\left( s-\widetilde{m}_Q^2\right)\nonumber \\
&&+\frac{m_Q^2\langle\bar{s} s\rangle^2}{72 T^6}\langle\frac{\alpha_sGG}{\pi}\rangle\int_0^1 dy \, \widetilde{m}_Q^4 \, \delta \left( s-\widetilde{m}_Q^2\right) \, ,
\end{eqnarray}
the subscripts  $0$, $3$, $4$, $5$, $6$, $7$, $8$, $10$ denote the dimensions of the  vacuum condensates, $y_{f}=\frac{1+\sqrt{1-4m_Q^2/s}}{2}$,
$y_{i}=\frac{1-\sqrt{1-4m_Q^2/s}}{2}$, $z_{i}=\frac{y
m_Q^2}{y s -m_Q^2}$, $\overline{m}_Q^2=\frac{(y+z)m_Q^2}{yz}$,
$ \widetilde{m}_Q^2=\frac{m_Q^2}{y(1-y)}$, $\int_{y_i}^{y_f}dy \to \int_{0}^{1}dy$, $\int_{z_i}^{1-y}dz \to \int_{0}^{1-y}dz$ when the $\delta$ functions $\delta\left(s-\overline{m}_Q^2\right)$ and $\delta\left(s-\widetilde{m}_Q^2\right)$ appear.
 In this article, we carry out the
operator product expansion to the vacuum condensates  up to dimension-10, and  assume  vacuum saturation for the  higher dimensional vacuum condensates. The condensates $\langle \frac{\alpha_s}{\pi}GG\rangle$, $\langle \bar{s}s\rangle\langle \frac{\alpha_s}{\pi}GG\rangle$,
$\langle \bar{s}s\rangle^2\langle \frac{\alpha_s}{\pi}GG\rangle$, $\langle \bar{s} g_s \sigma Gs\rangle^2$ and $g_s^2\langle \bar{s}s\rangle^2$ are the vacuum expectations
of the operators of the order
$\mathcal{O}(\alpha_s)$.  The four-quark condensate $g_s^2\langle \bar{q}q\rangle^2$ comes from the terms
$\langle \bar{s}\gamma_\mu t^a s g_s D_\eta G^a_{\lambda\tau}\rangle$, $\langle\bar{s}_jD^{\dagger}_{\mu}D^{\dagger}_{\nu}D^{\dagger}_{\alpha}s_i\rangle$  and
$\langle\bar{s}_jD_{\mu}D_{\nu}D_{\alpha}s_i\rangle$, rather than comes from the perturbative corrections of $\langle \bar{s}s\rangle^2$.
 The condensates $\langle g_s^3 GGG\rangle$, $\langle \frac{\alpha_s GG}{\pi}\rangle^2$,
 $\langle \frac{\alpha_s GG}{\pi}\rangle\langle \bar{s} g_s \sigma Gs\rangle$ have the dimensions 6, 8, 9 respectively,  but they are   the vacuum expectations
of the operators of the order    $\mathcal{O}( \alpha_s^{3/2})$, $\mathcal{O}(\alpha_s^2)$, $\mathcal{O}( \alpha_s^{3/2})$ respectively, and discarded.  We take
the truncations $n\leq 10$ and $k\leq 1$ in a consistent way,
the operators of the orders $\mathcal{O}( \alpha_s^{k})$ with $k> 1$ are  discarded.
 Furthermore,  the values of the  condensates $\langle g_s^3 GGG\rangle$, $\langle \frac{\alpha_s GG}{\pi}\rangle^2$,
 $\langle \frac{\alpha_s GG}{\pi}\rangle\langle \bar{s} g_s \sigma Gs\rangle$   are very small, and they can be  neglected safely.
In Refs.\cite{WangHuang-molecule,WangHuangTao1312,Wang1311,WangHuangTao}, the same truncations are taken to study the hidden-charmed and hidden-bottom tetraquark states and molecular states with the QCD sum rules, and to obtain the energy scale formula, such truncations work well.

 Differentiate   Eq.(22) with respect to  $\frac{1}{T^2}$, then eliminate the
 pole residues $\lambda_{Y/Z}$, we obtain the QCD sum rules for
 the masses of the scalar, axial-vector and tensor    $D_s^*\bar{D}_s^*$ and   $B_s^*\bar{B}_s^*$ molecular states,
 \begin{eqnarray}
 M^2_{Y/Z}= \frac{\int_{4m_Q^2}^{s_0} ds\frac{d}{d \left(-1/T^2\right)}\rho(s)\exp\left(-\frac{s}{T^2}\right)}{\int_{4m_Q^2}^{s_0} ds \rho(s)\exp\left(-\frac{s}{T^2}\right)}\, .
\end{eqnarray}

We can obtain the QCD sum rules for the  $D^*\bar{D}^*$ and   $B^*\bar{B}^*$ molecular states with the simple replacements,
 \begin{eqnarray}
m_s &\to& 0\, , \nonumber\\
\langle\bar{s}s\rangle &\to& \langle\bar{q}q\rangle\, , \nonumber\\
\langle\bar{s}g_s\sigma G s\rangle &\to& \langle\bar{q}g_s\sigma Gq\rangle\, .
\end{eqnarray}

For the tetraquark and molecular states, it is more reasonable to refer to the $\lambda_{X/Y/Z}$ as the pole residues (not the decay constants).
We cannot obtain the true values of the pole residues $\lambda_{X/Y/Z}$ by measuring the leptonic decays as in the cases of the $D_{s}(D)$ and $J/\psi (\Upsilon)$,
$D_{s}(D)\to \ell\nu$ and $J/\psi (\Upsilon)\to e^+e^-$, and have to calculate the  $\lambda_{X/Y/Z}$ using  some theoretical methods. It is hard to obtain the true values. In this article, we focus on the masses to study the molecular states, and the unknown contributions of the perturbative corrections to the pole residues in the numerator and denominator are expected to be canceled out with each other efficiently,  as we obtain the hadronic masses $M_{Y/Z}$ through a ratio, see Eq.(46). Neglecting perturbative ${\mathcal{O}}(\alpha_s)$ corrections cannot impair the predictive ability qualitatively.

\section{Numerical results and discussions}
The vacuum condensates are taken to be the standard values
$\langle\bar{q}q \rangle=-(0.24\pm 0.01\, \rm{GeV})^3$, $\langle\bar{s}s \rangle=(0.8\pm0.1)\langle\bar{q}q \rangle$,
$\langle\bar{q}g_s\sigma G q \rangle=m_0^2\langle \bar{q}q \rangle$,
$\langle\bar{s}g_s\sigma G s \rangle=m_0^2\langle \bar{s}s \rangle$,
$m_0^2=(0.8 \pm 0.1)\,\rm{GeV}^2$, $\langle \frac{\alpha_s
GG}{\pi}\rangle=(0.33\,\rm{GeV})^4 $    at the energy scale  $\mu=1\, \rm{GeV}$
\cite{SVZ79,Reinders85,Ioffe2005}.
The quark condensates and mixed quark condensates evolve with the   renormalization group equation,
$\langle\bar{q}q \rangle(\mu)=\langle\bar{q}q \rangle(Q)\left[\frac{\alpha_{s}(Q)}{\alpha_{s}(\mu)}\right]^{\frac{4}{9}}$,
$\langle\bar{s}s \rangle(\mu)=\langle\bar{s}s \rangle(Q)\left[\frac{\alpha_{s}(Q)}{\alpha_{s}(\mu)}\right]^{\frac{4}{9}}$,
 $\langle\bar{q}g_s \sigma Gq \rangle(\mu)=\langle\bar{q}g_s \sigma Gq \rangle(Q)\left[\frac{\alpha_{s}(Q)}{\alpha_{s}(\mu)}\right]^{\frac{2}{27}}$,
 $\langle\bar{s}g_s \sigma Gs \rangle(\mu)=\langle\bar{s}g_s \sigma Gs \rangle(Q)\left[\frac{\alpha_{s}(Q)}{\alpha_{s}(\mu)}\right]^{\frac{2}{27}}$.

In the article, we take the $\overline{MS}$ masses $m_{c}(m_c)=(1.275\pm0.025)\,\rm{GeV}$, $m_{b}(m_b)=(4.18\pm0.03)\,\rm{GeV}$ and $m_s(\mu=2\,\rm{GeV})=(0.095\pm0.005)\,\rm{GeV}$
 from the Particle Data Group \cite{PDG}, and take into account
the energy-scale dependence of  the $\overline{MS}$ masses from the renormalization group equation,
\begin{eqnarray}
m_s(\mu)&=&m_s({\rm 2GeV} )\left[\frac{\alpha_{s}(\mu)}{\alpha_{s}({\rm 2GeV})}\right]^{\frac{4}{9}} \, ,\nonumber\\
m_c(\mu)&=&m_c(m_c)\left[\frac{\alpha_{s}(\mu)}{\alpha_{s}(m_c)}\right]^{\frac{12}{25}} \, ,\nonumber\\
m_b(\mu)&=&m_b(m_b)\left[\frac{\alpha_{s}(\mu)}{\alpha_{s}(m_b)}\right]^{\frac{12}{23}} \, ,\nonumber\\
\alpha_s(\mu)&=&\frac{1}{b_0t}\left[1-\frac{b_1}{b_0^2}\frac{\log t}{t} +\frac{b_1^2(\log^2{t}-\log{t}-1)+b_0b_2}{b_0^4t^2}\right]\, ,
\end{eqnarray}
  where $t=\log \frac{\mu^2}{\Lambda^2}$, $b_0=\frac{33-2n_f}{12\pi}$, $b_1=\frac{153-19n_f}{24\pi^2}$, $b_2=\frac{2857-\frac{5033}{9}n_f+\frac{325}{27}n_f^2}{128\pi^3}$,  $\Lambda=213\,\rm{MeV}$, $296\,\rm{MeV}$  and  $339\,\rm{MeV}$ for the flavors  $n_f=5$, $4$ and $3$, respectively  \cite{PDG}.

In the conventional QCD sum rules \cite{SVZ79,Reinders85}, there are
two criteria (pole dominance and convergence of the operator product
expansion) for choosing  the Borel parameter $T^2$ and threshold
parameter $s_0$.  We impose
the two criteria on the hidden charmed (or bottom) molecular states, and search for the optimal  values.

In Refs.\cite{WangHuang-molecule,WangHuangTao1312,Wang1311,WangHuangTao}, we study the acceptable energy scales of the QCD spectral densities in the QCD sum rules for the hidden  charmed (bottom) tetraquark states and molecular states in details for the first time,  and suggest a  formula $\mu=\sqrt{M^2_{X/Y/Z}-(2{\mathbb{M}}_Q)^2}$ to determine  the energy scales of the QCD spectral densities.
The heavy tetraquark system $Q\bar{Q}q^{\prime}\bar{q}$ could be described
by a double-well potential with two light quarks $q^{\prime}\bar{q}$ lying in the two wells respectively.
   In the heavy quark limit, the $Q$-quark can be taken as a static well potential,
which binds the light quark $q^{\prime}$ to form a diquark in the color antitriplet channel or binds the light antiquark $\bar{q}$ to
form a meson in the color singlet channel (or a meson-like state in the color octet  channel).
Then the heavy tetraquark states  are characterized by the effective heavy quark masses ${\mathbb{M}}_Q$ (or constituent quark masses) and
the virtuality $V=\sqrt{M^2_{X/Y/Z}-(2{\mathbb{M}}_Q)^2}$ (or bound energy not as robust). The effective masses ${\mathbb{M}}_Q$, just like the mixed condensates,
 appear as parameters and their values are fitted by the QCD sum rules.
The effective masses  ${\mathbb{M}}_Q$ have uncertainties, the optimal values in the diquark-antidiquark systems are not necessary the ideal values in the
 meson-meson systems.  The QCD sum rules have three typical energy scales $\mu^2$, $T^2$, $V^2$.
 It is natural to take the energy  scale,
$ \mu^2=V^2={\mathcal{O}}(T^2)$.
 The effective masses ${\mathbb{M}}_c=1.84\,\rm{GeV}$ and ${\mathbb{M}}_b=5.14\,\rm{GeV}$ are the optimal values for  the hadronic molecular states, and
 can reproduce the experimental data $M_{X(3872)}=3.87\,\rm{GeV}$, $M_{Z_c(3900)}=3.90\,\rm{GeV}$, $M_{Z_b(10610)}=10.61\,\rm{GeV}$ approximately \cite{WangHuang-molecule}. In this article, we take the effective masses ${\mathbb{M}}_c=1.84\,\rm{GeV}$ and ${\mathbb{M}}_b=5.14\,\rm{GeV}$, and the predictions indicate that they are also the optimal values to reproduce the experimental values of the masses of the $Z_c(4020)$, $Z_c(4025)$, $Y(4140)$ and $Z_b(10650)$.

The energy scale formula serves as additional constraints  on choosing the Borel parameters and threshold parameters, as the predicted masses should satisfy the formula.  The optimal Borel parameters and continuum threshold parameters therefore  the pole contributions and energy scales of the QCD spectral densities are shown explicitly in Table 1.

In Fig.1,  the masses of the scalar $D^*\bar{D}^*$ and  $D_s^*\bar{D}_s^*$ molecular states are plotted   with variations of the  Borel parameters $T^2$ and energy scales $\mu$ for the continuum threshold parameters  $s^0_{D^*\bar{D}^*}=20\,\rm{GeV}^2$ and $s^0_{D_s^*\bar{D}_s^*}=22\,\rm{GeV}^2$, respectively. From the figure, we can see that the masses decrease monotonously with increase of the energy scales, the energy scales $\mu = (1.5-1.6)\,\rm{GeV}$ and $\mu = (1.7-1.9)\,\rm{GeV}$ can reproduce the experimental values of the masses $M_{Z_c(4025)}$ (or $M_{Z_c(4020)}$) and $M_{Y(4140)}$, respectively. The formula $\mu=\sqrt{M^2_{X/Y/Z}-(2{\mathbb{M}}_Q)^2}$ leads  to  the values $\mu=1.6\,\rm{GeV}$ and $\mu=1.8\,\rm{GeV}$ for the scalar  $D^*\bar{D}^*$ and  $D_s^*\bar{D}_s^*$ molecular states, respectively. The agreements are excellent. The masses $M_{Y(3940)}< M_{Z_c(4025)}$, the energy scale of the QCD spectral density of the $Y(3940)$ should be smaller than that of the $Z_c(4025)$ according to the energy formula. From Fig.1, we can see that the predicted mass is larger than $3.95\,\rm{GeV}$ even for the energy scale $\mu=1.8\,\rm{GeV}$, and we cannot satisfy the relation $\sqrt{s_0}\approx M_{Y(3940)}+0.5\,\rm{GeV}$ with reasonable  $M_{Y(3940)}$ compared to the experimental data. Now the $X(3915)$ is listed in the Review of Particle Physics as the $\chi_{c0}({\rm 2P})$
state with $J^{PC}=0^{++}$ \cite{PDG}. The present result  supports the assignment of the Particle Data Group.
In Ref.\cite{Y4140-Wang}, we study the scalar $D^*\bar{D}^*$,  $D_s^*\bar{D}_s^*$, $B^*\bar{B}^*$,  $B_s^*\bar{B}_s^*$ molecular states with the QCD sum rules by  carrying  out the operator product expansion to the vacuum condensates  up to
dimension-10 and setting the energy scale to be $\mu=1\,\rm{GeV}$. The predicted masses
 are about $(250-500)\,\rm{MeV}$ above the
corresponding ${D}^\ast {\bar {D}}^\ast$, ${D}_s^\ast {\bar
{D}}_s^\ast$, ${B}^\ast{\bar {B}}^\ast$ and ${B}_s^\ast {\bar
{B}}_s^\ast$ thresholds. If larger energy scales are taken, the conclusion should be modified.

In Figs.2-3,  the contributions of different terms in the
operator product expansion are plotted with variations of the Borel parameters  $T^2$ for the energy scales and central values of the  threshold parameters shown  in Table 1.  The contributions of the condensates do not decrease monotonously with increase of dimensions. However, in the Borel windows shown in Table 1, the $D_4$, $D_7$, $D_{10}$ play a less important role,
$D_3 \gg |D_5|\gg D_6\gg |D_8|$ for the $J=2$ molecular states and $J=0$ $D_s^*\bar{D}_s^*$ molecular state,
$D_3 \gg |D_5|\gg D_6$ for the $J=1$ molecular states,
$D_3 \gg |D_5|\sim D_6\gg |D_8|$ for the $J=0$ $D^*\bar{D}^*$ and $B_s^*\bar{B}_s^*$ molecular states,
$D_3 > D_6 > |D_5|\sim |D_8|$ for the $J=0$ $B^*\bar{B}^*$ molecular state,    the $D_6$,
$D_8$, $D_{10}$ decrease monotonously and quickly  with increase of the Borel parameters for the $J=0,2$ molecular states,
where the $D_i$ with $i=0,\,3,\,4,\,5,\,6,\,7,\,8,\,10$ denote the contributions of the vacuum condensates of dimensions $D=i$, and the total contributions are normalized to be $1$.
The convergence of the operator product expansion does not mean that the perturbative terms make dominant contributions,
as the  continuum hadronic spectral densities  are approximated by
$\rho_{QCD}(s)\Theta(s-s_0)$ in the QCD sum rules for the heavy molecular states, where
the $\rho_{QCD}(s)$ denotes the full QCD spectral densities; the contributions of the quark condensates $\langle\bar{q}q\rangle$ and $\langle\bar{s}s\rangle$ (of dimension-3) can be very large.
  In summary, the two criteria (pole dominance and convergence of the operator product expansion) of the QCD sum rules are fully satisfied, so we expect to make reasonable predictions.

\begin{table}
\begin{center}
\begin{tabular}{|c|c|c|c|c|c|c|c|c|}\hline\hline
$J^{PC}$                     &$\mu(\rm{GeV})$&$T^2 (\rm{GeV}^2)$ &$s_0 (\rm{GeV}^2)$&pole        &$M_{Y/Z}(\rm{GeV})$       &$\lambda_{Y/Z}(\rm{GeV}^{5(4)})$ \\ \hline

$0^{++}$ ($c\bar{c}u\bar{d}$)&1.6            &$2.5-2.9$          &$20\pm1$          &$(43-68)\%$ &$4.01^{+0.09}_{-0.09}$  &$3.97^{+0.67}_{-0.60}\times10^{-2}$ \\ \hline
$1^{+-}$ ($c\bar{c}u\bar{d}$)&1.7            &$2.8-3.2$          &$20\pm1$          &$(45-68)\%$ &$4.04^{+0.07}_{-0.08}$  &$6.37^{+0.96}_{-0.89}\times10^{-3}$ \\ \hline
$2^{++}$ ($c\bar{c}u\bar{d}$)&1.6            &$2.6-3.0$          &$20\pm1$          &$(45-69)\%$ &$4.01^{+0.10}_{-0.08}$  &$3.05^{+0.47}_{-0.44}\times10^{-2}$ \\ \hline

$0^{++}$ ($c\bar{c}s\bar{s}$)&1.8            &$2.8-3.2$          &$22\pm1$          &$(46-69)\%$ &$4.14^{+0.08}_{-0.08}$  &$5.75^{+0.96}_{-0.85}\times10^{-2}$ \\ \hline
$1^{+-}$ ($c\bar{c}s\bar{s}$)&1.9            &$3.2-3.6$          &$22\pm1$          &$(48-68)\%$ &$4.16^{+0.05}_{-0.04}$  &$8.80^{+0.60}_{-0.57}\times10^{-3}$ \\ \hline
$2^{++}$ ($c\bar{c}s\bar{s}$)&1.8            &$3.0-3.4$          &$22\pm1$          &$(47-68)\%$ &$4.13^{+0.08}_{-0.08}$  &$4.34^{+0.67}_{-0.60}\times10^{-2}$ \\ \hline

$0^{++}$ ($b\bar{b}u\bar{d}$)&2.8            &$6.8-7.8$          &$124\pm2$         &$(44-65)\%$ &$10.65^{+0.15}_{-0.09}$ &$2.07^{+0.45}_{-0.32}\times10^{-1}$ \\ \hline
$1^{+-}$ ($b\bar{b}u\bar{d}$)&2.9            &$7.0-8.0$          &$124\pm2$         &$(45-65)\%$ &$10.67^{+0.09}_{-0.08}$ &$1.34^{+0.20}_{-0.18}\times10^{-2}$ \\ \hline
$2^{++}$ ($b\bar{b}u\bar{d}$)&2.8            &$7.2-8.2$          &$124\pm2$         &$(44-64)\%$ &$10.66^{+0.14}_{-0.09}$ &$1.67^{+0.31}_{-0.23}\times10^{-1}$ \\ \hline

$0^{++}$ ($b\bar{b}s\bar{s}$)&2.9            &$7.0-8.0$          &$126\pm2$         &$(45-66)\%$ &$10.70^{+0.11}_{-0.08}$ &$2.49^{+0.41}_{-0.35}\times10^{-1}$ \\ \hline
$1^{+-}$ ($b\bar{b}s\bar{s}$)&3.0            &$7.2-8.2$          &$126\pm2$         &$(47-66)\%$ &$10.73^{+0.09}_{-0.07}$ &$1.63^{+0.23}_{-0.21}\times10^{-2}$ \\ \hline
$2^{++}$ ($b\bar{b}s\bar{s}$)&3.0            &$7.8-8.8$          &$128\pm2$         &$(48-66)\%$ &$10.71^{+0.08}_{-0.08}$ &$2.31^{+0.31}_{-0.27}\times10^{-1}$ \\ \hline
 \hline
\end{tabular}
\end{center}
\caption{ The Borel parameters, continuum threshold parameters, pole contributions, energy scales,  masses and pole residues of the scalar, axial-vector and tensor  molecular states. The symbolic quark constituents are shown in the bracket.   }
\end{table}

\begin{figure}
\centering
\includegraphics[totalheight=6cm,width=7cm]{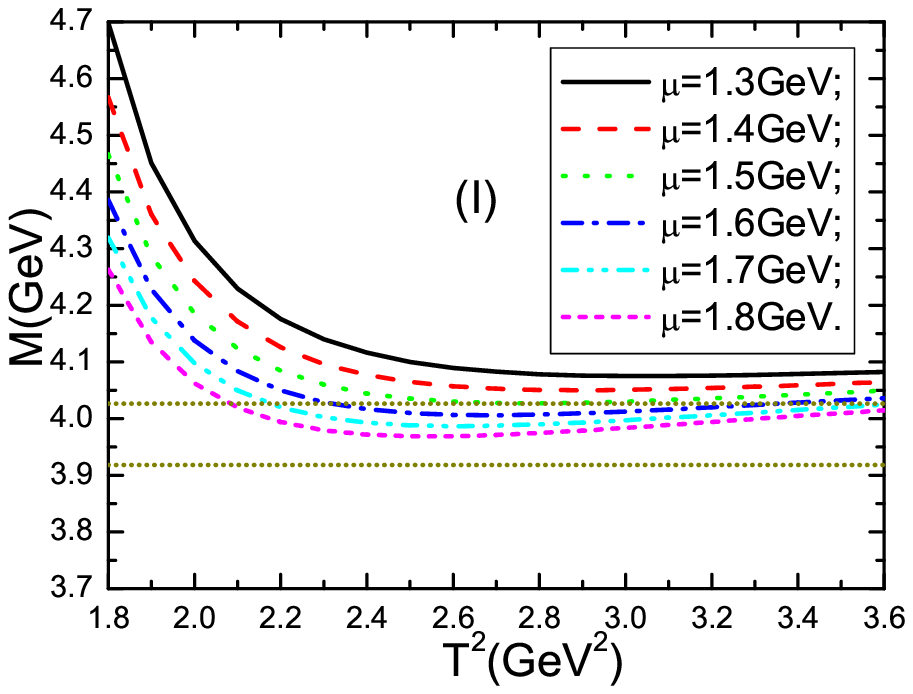}
\includegraphics[totalheight=6cm,width=7cm]{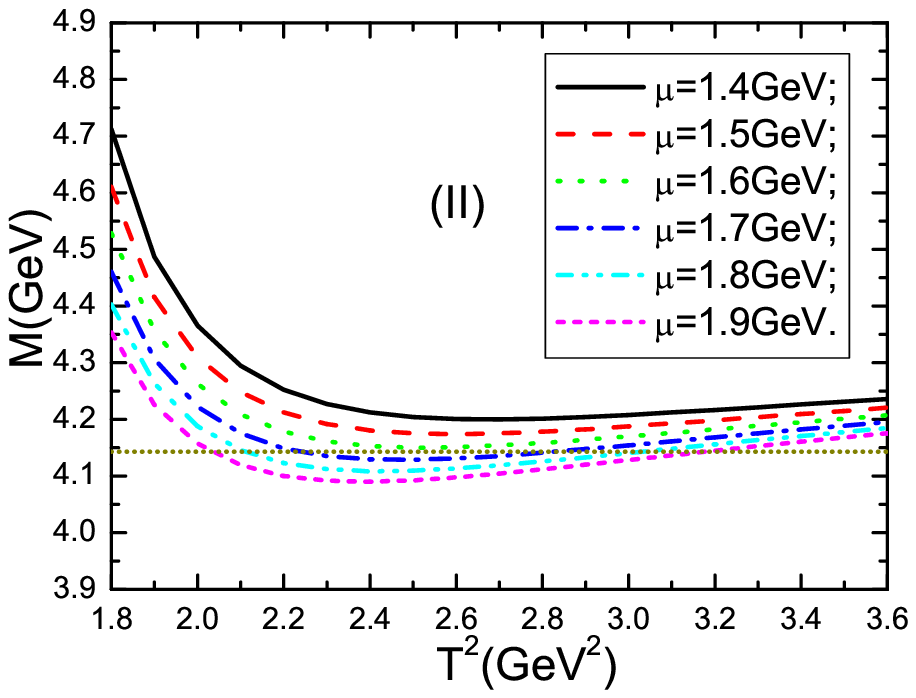}
  \caption{ The masses  with variations of the  Borel parameters $T^2$ and energy scales $\mu$, where the horizontal lines denote the experimental values of the masses of the $Z_c(4025)$, $Y(3940)$ and $Y(4140)$, respectively, the (I) and (II) denote the scalar $D^*\bar{D}^*$ and $D_s^*\bar{D}_s^*$ molecular states, respectively. }
\end{figure}

\begin{figure}
\centering
\includegraphics[totalheight=6cm,width=7cm]{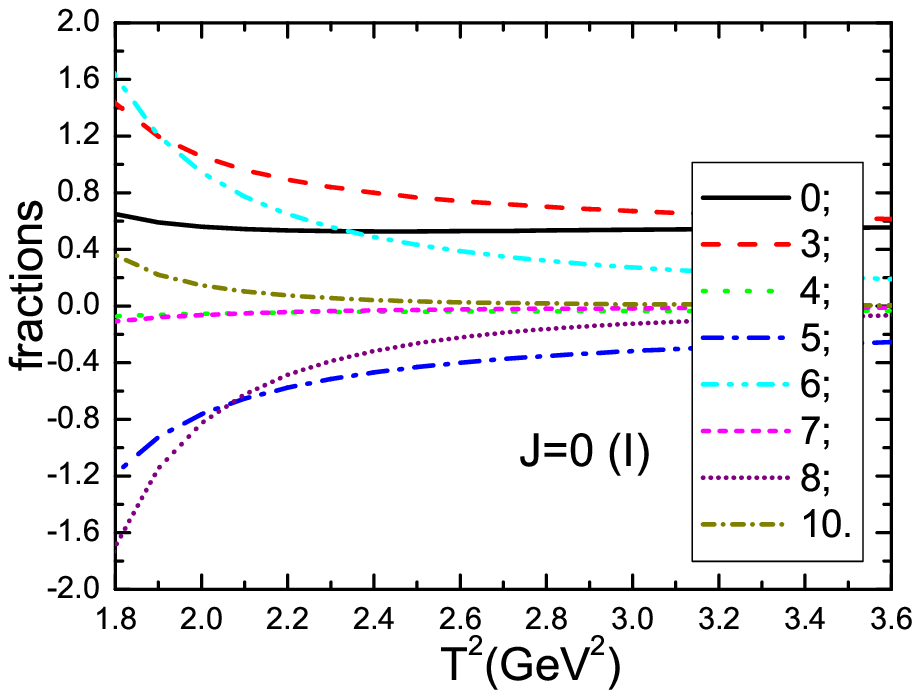}
\includegraphics[totalheight=6cm,width=7cm]{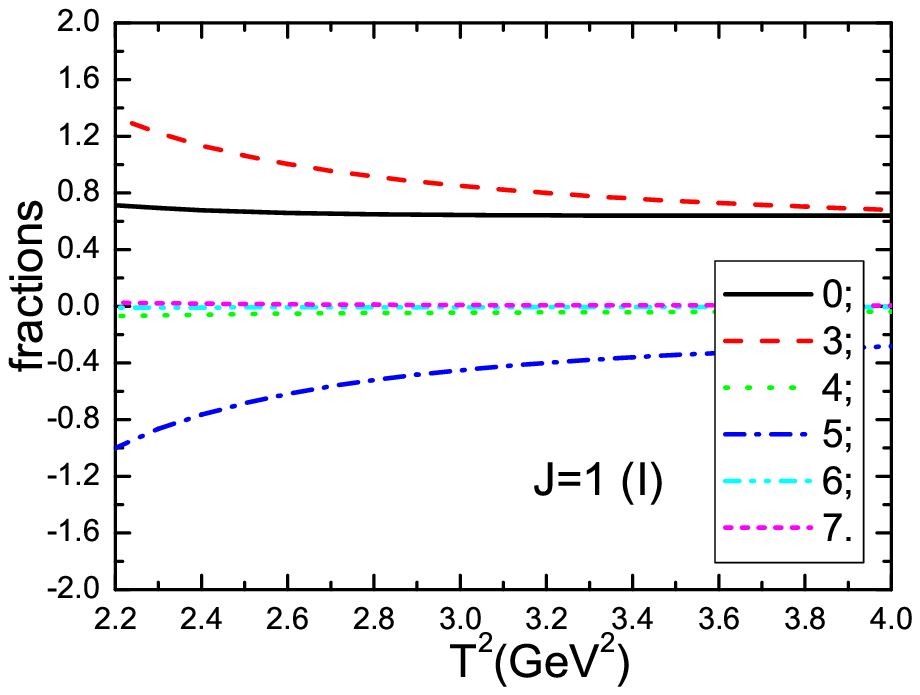}
\includegraphics[totalheight=6cm,width=7cm]{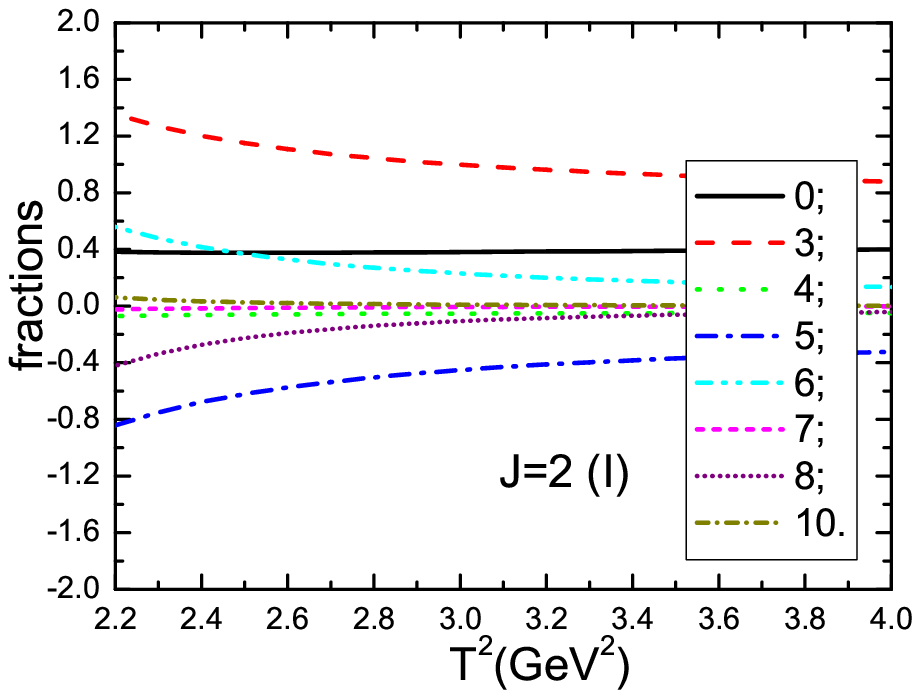}
\includegraphics[totalheight=6cm,width=7cm]{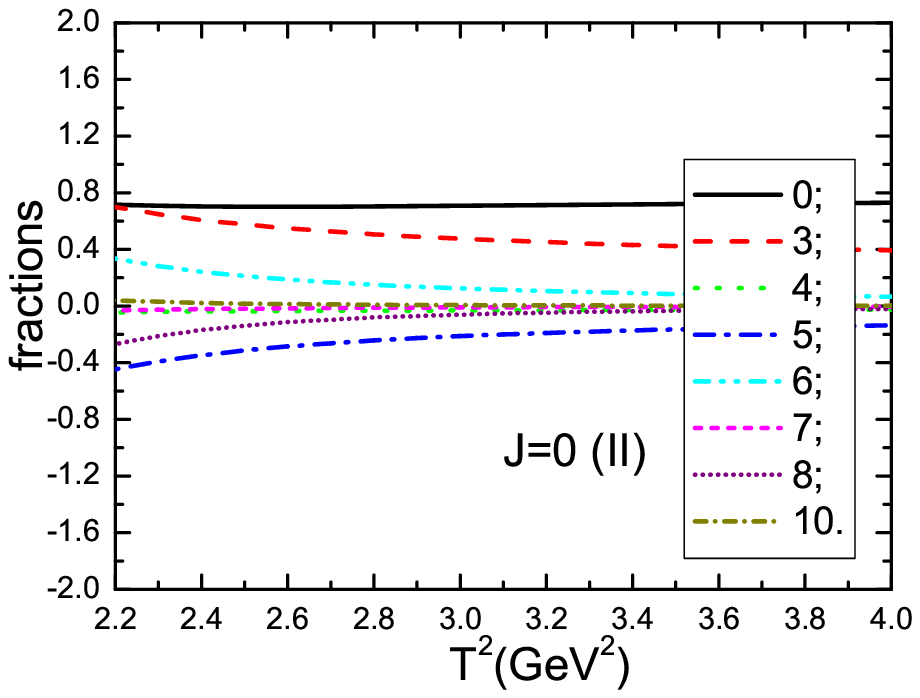}
\includegraphics[totalheight=6cm,width=7cm]{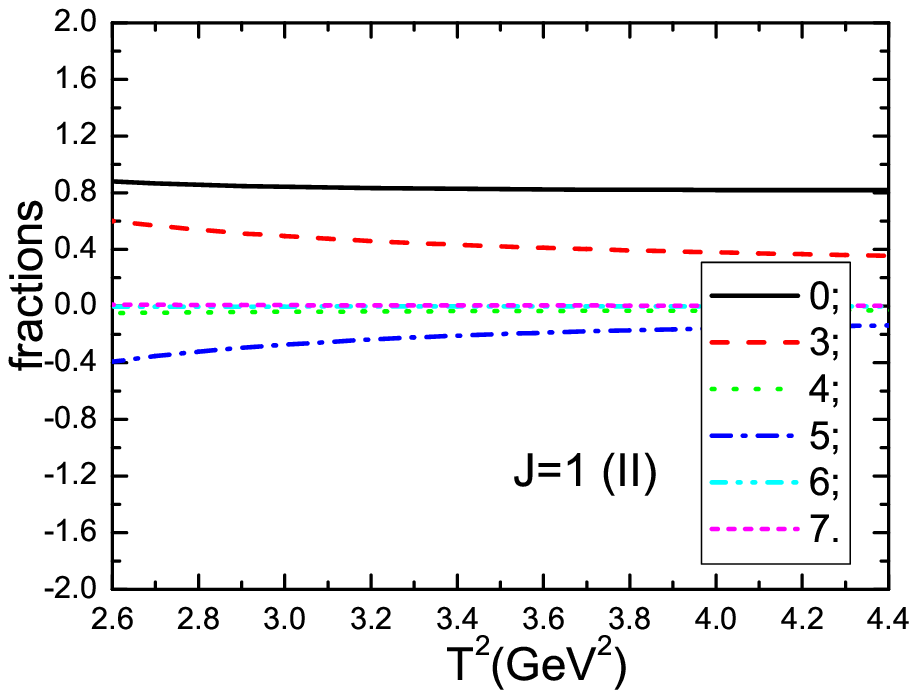}
\includegraphics[totalheight=6cm,width=7cm]{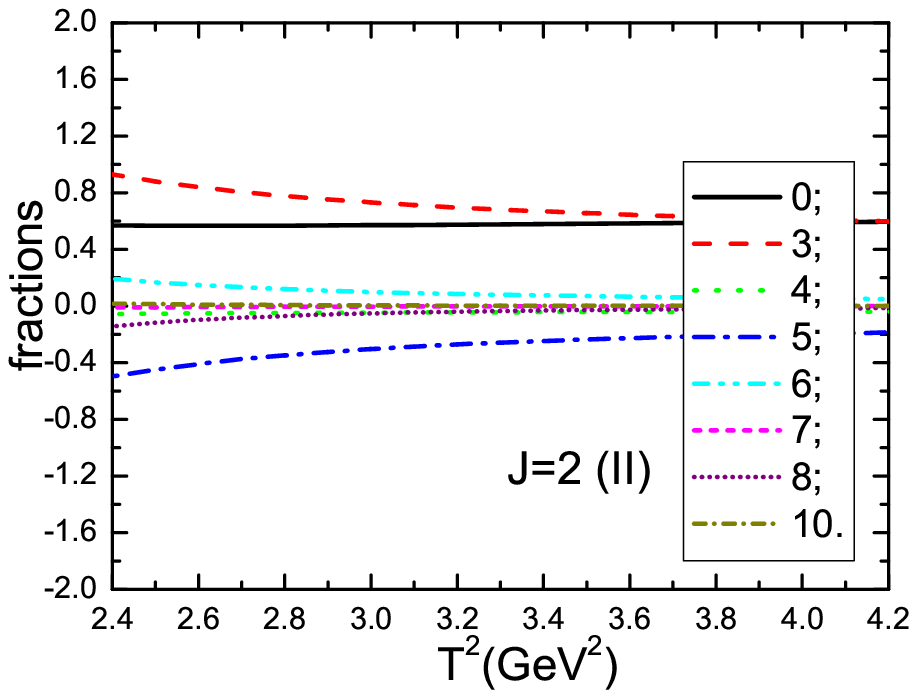}
  \caption{ The contributions of different terms in the operator product expansion  with variations of the
  Borel parameters $T^2$, where the 0, 3, 4, 5, 6, 7, 8,  10 denotes the dimensions of the vacuum condensates, the $J=0,1,2$ denote the angular momentum of the molecular states, the (I) and (II) denote the $D^*\bar{D}^*$ and $D_s^*\bar{D}_s^*$ molecular states, respectively.  }
\end{figure}

\begin{figure}
\centering
\includegraphics[totalheight=6cm,width=7cm]{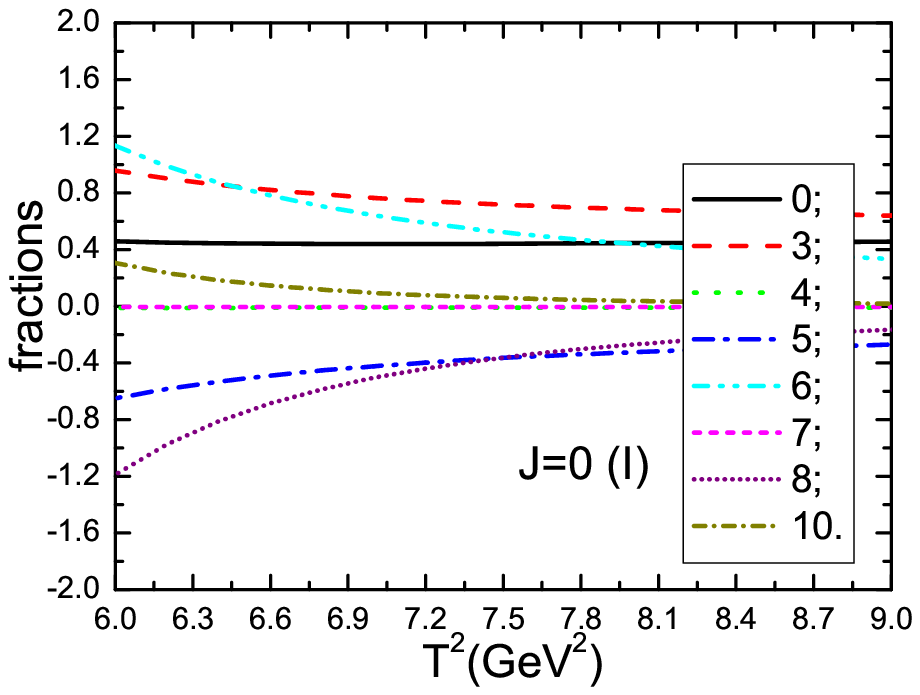}
\includegraphics[totalheight=6cm,width=7cm]{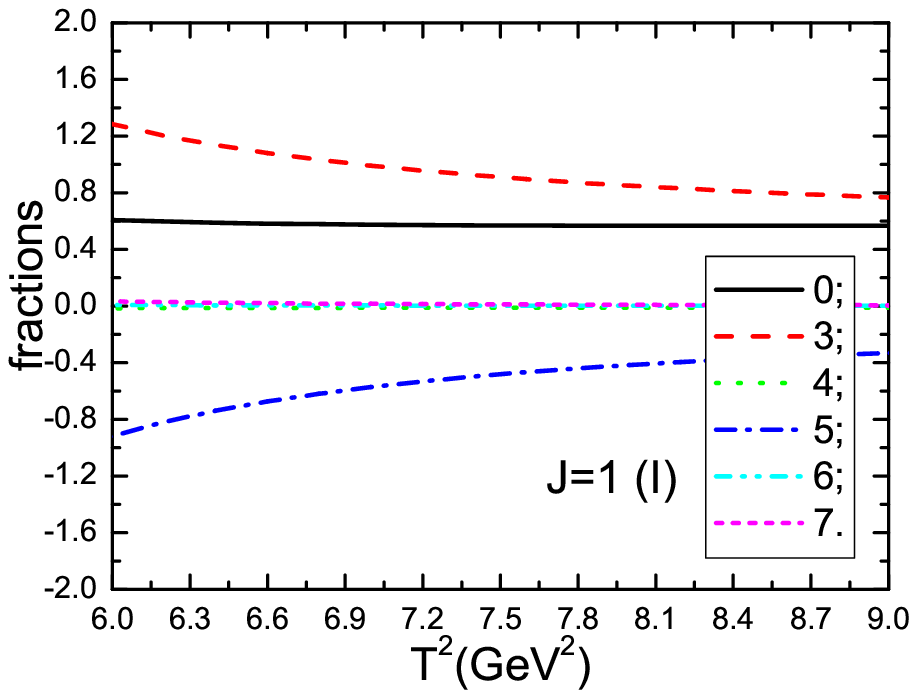}
\includegraphics[totalheight=6cm,width=7cm]{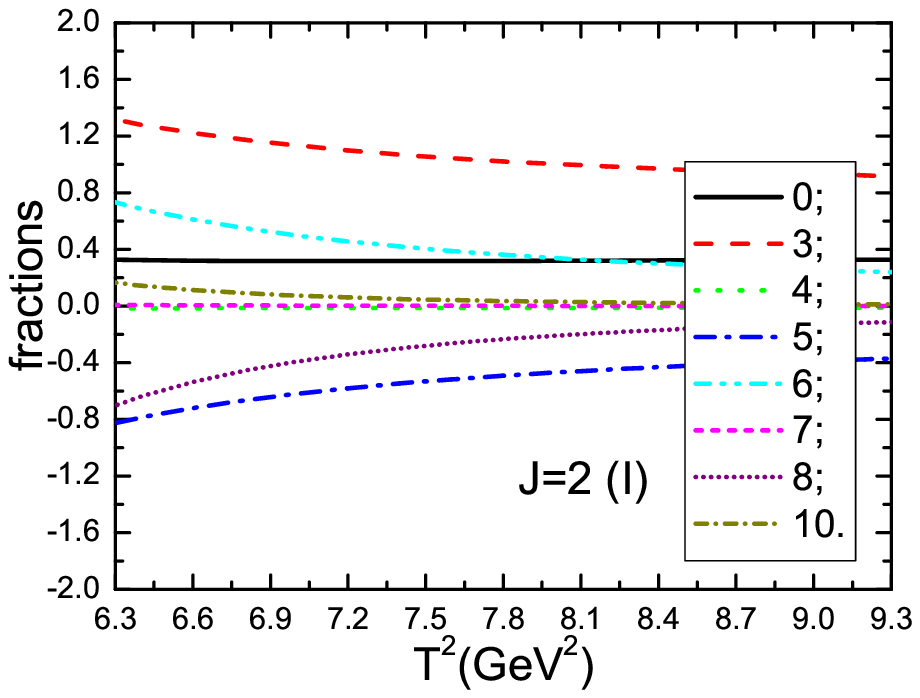}
\includegraphics[totalheight=6cm,width=7cm]{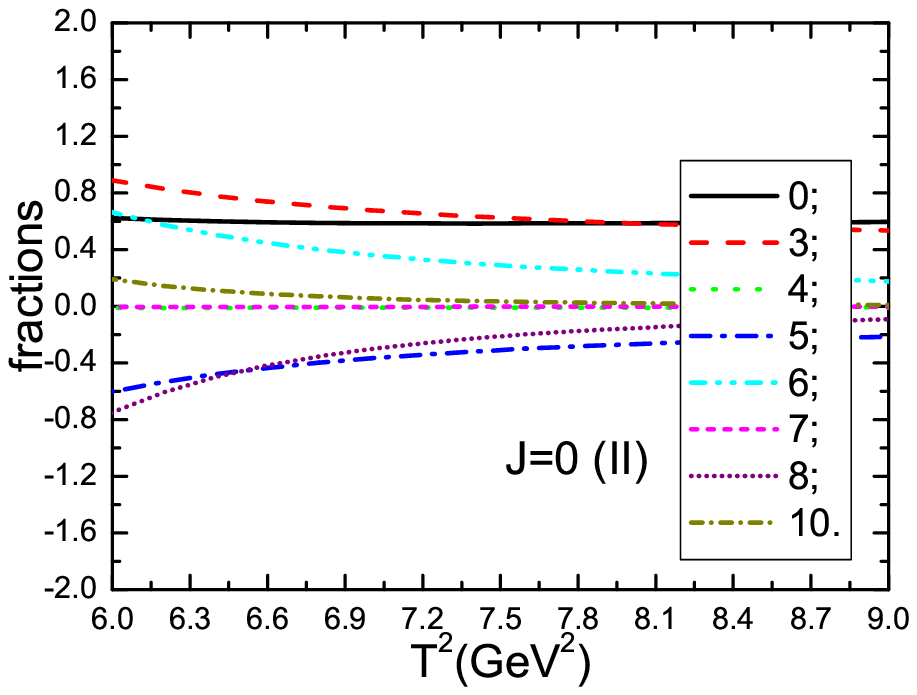}
\includegraphics[totalheight=6cm,width=7cm]{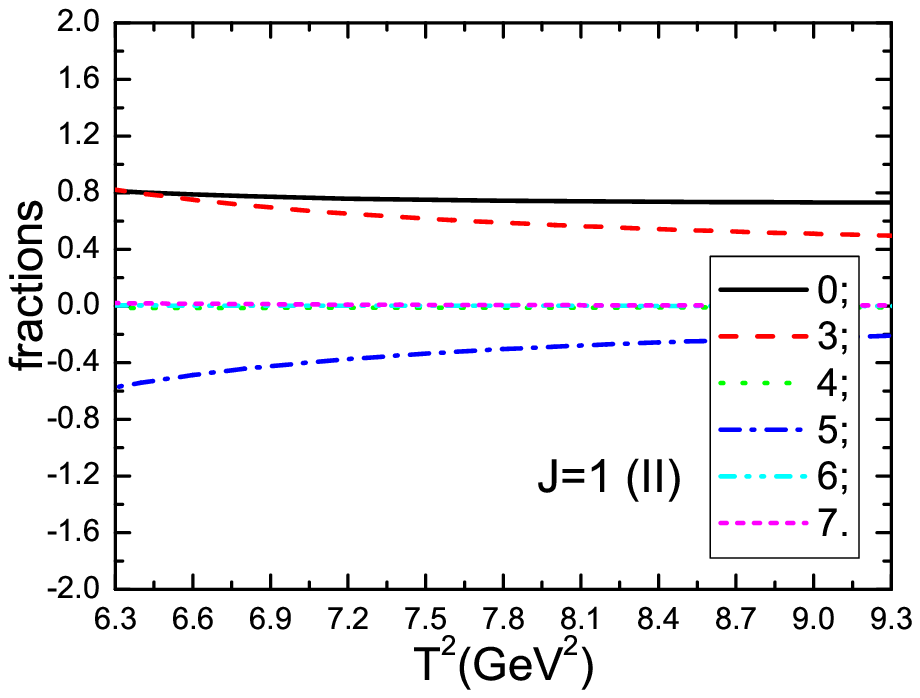}
\includegraphics[totalheight=6cm,width=7cm]{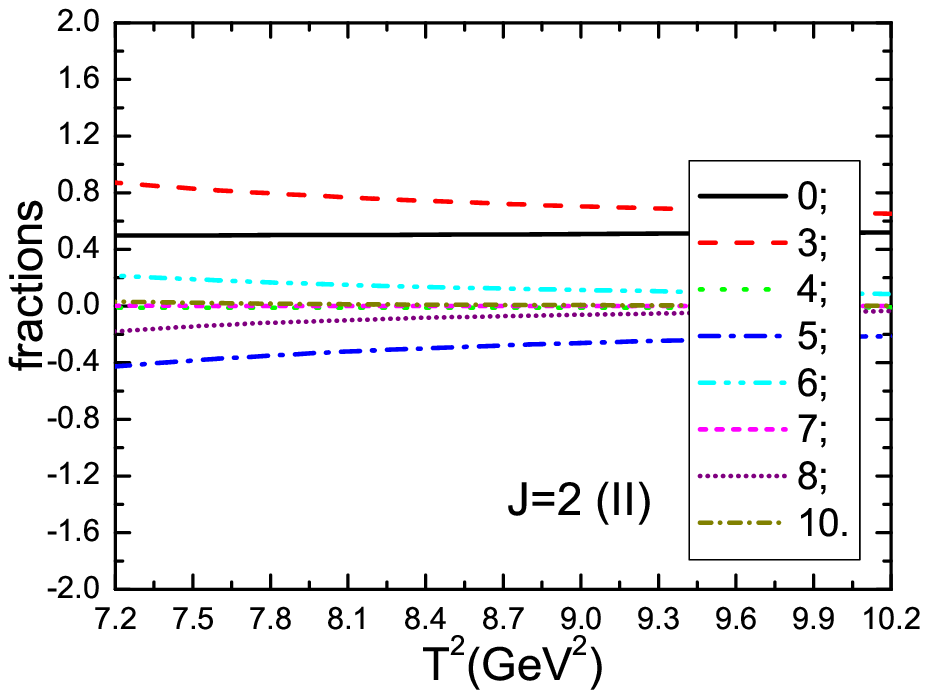}
  \caption{ The contributions of different terms in the operator product expansion  with variations of the
  Borel parameters $T^2$, where the 0, 3, 4, 5, 6, 7, 8,  10 denotes the dimensions of the vacuum condensates, the $J=0,1,2$ denote the angular momentum of the molecular states, the (I) and (II) denote the $B^*\bar{B}^*$ and $B_s^*\bar{B}_s^*$ molecular states, respectively. }
\end{figure}

\begin{figure}
\centering
\includegraphics[totalheight=6cm,width=7cm]{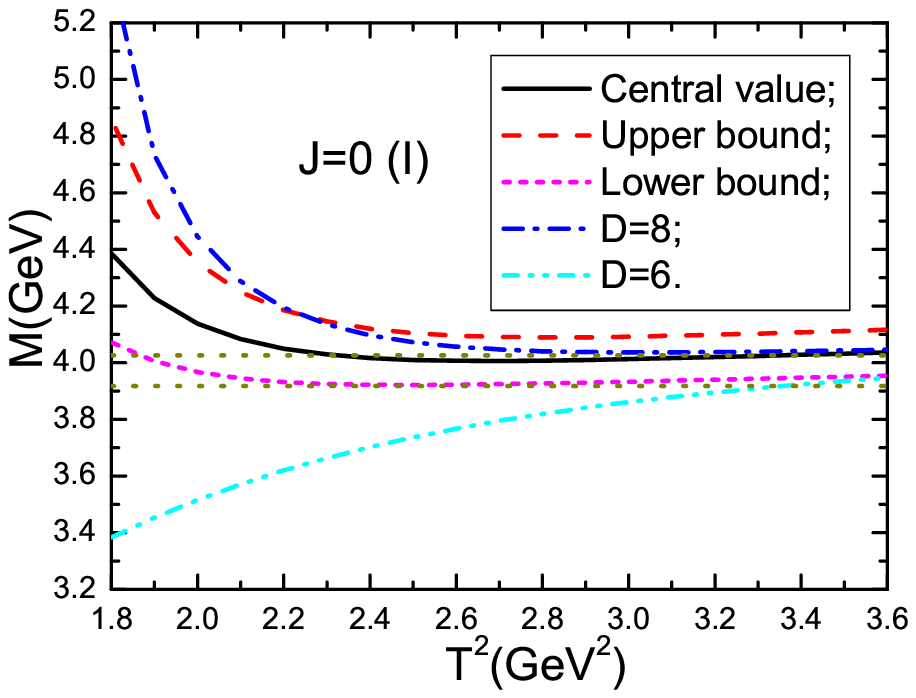}
\includegraphics[totalheight=6cm,width=7cm]{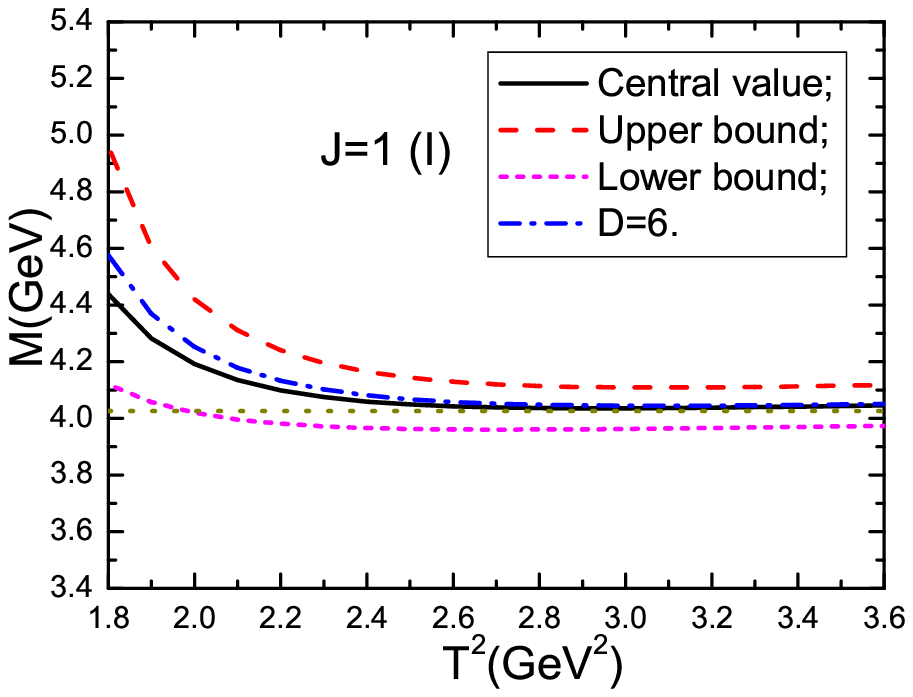}
\includegraphics[totalheight=6cm,width=7cm]{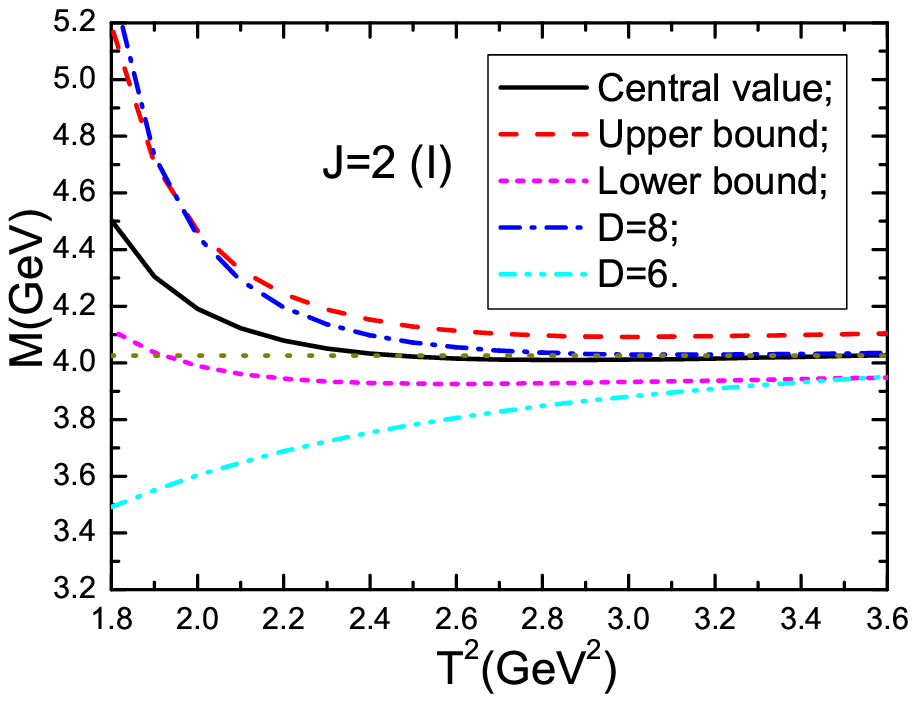}
\includegraphics[totalheight=6cm,width=7cm]{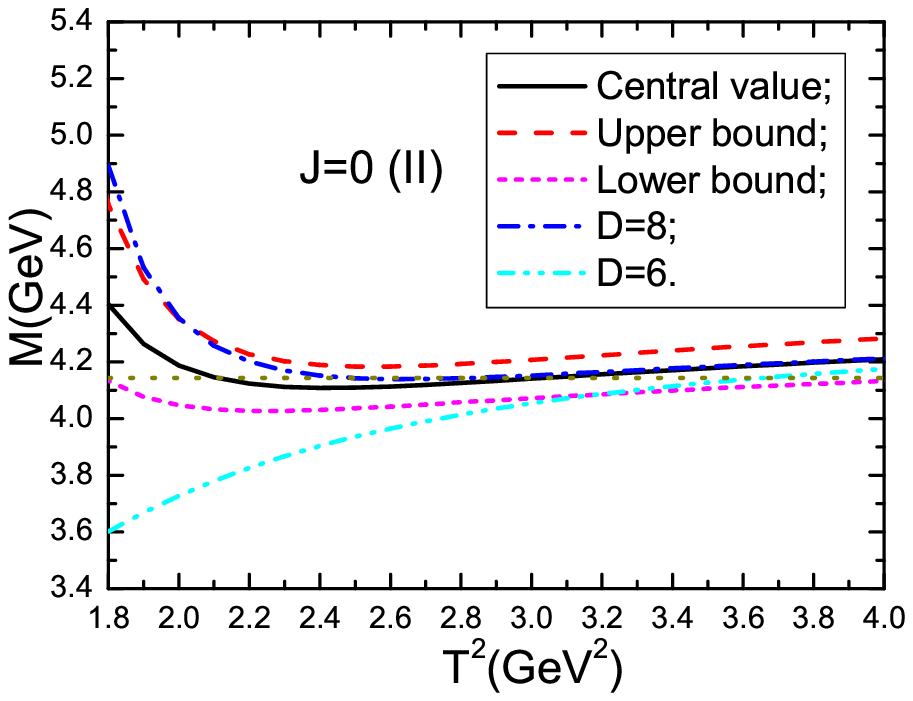}
\includegraphics[totalheight=6cm,width=7cm]{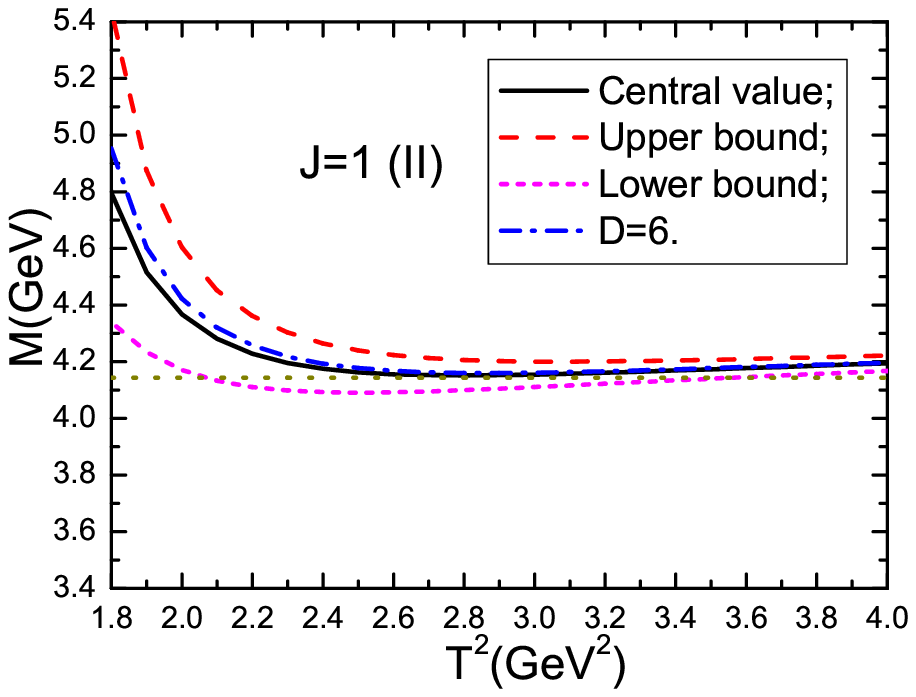}
\includegraphics[totalheight=6cm,width=7cm]{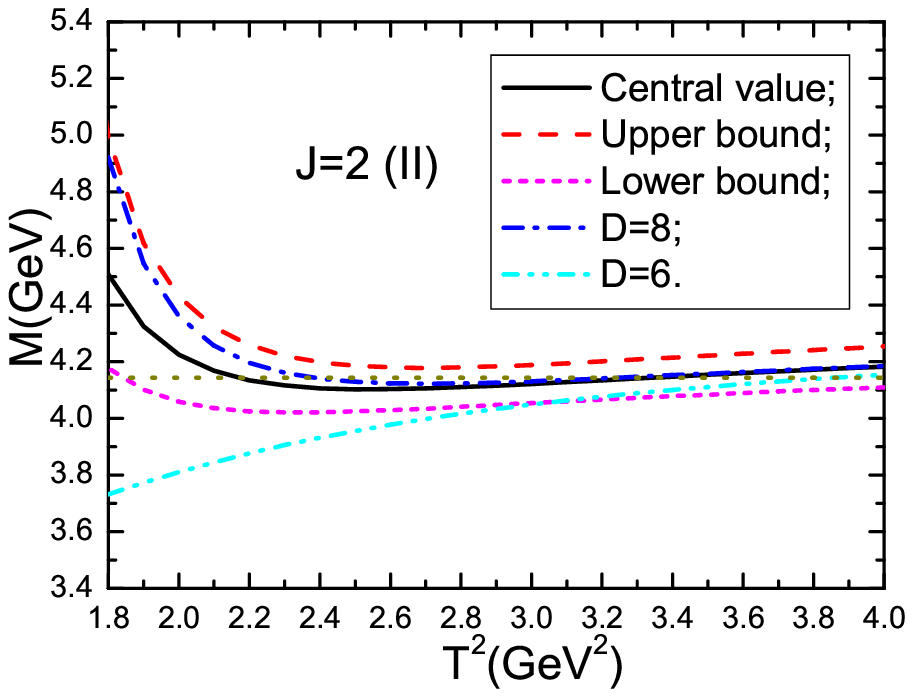}
  \caption{ The masses  with variations of the  Borel parameters $T^2$, where the horizontal lines denote the experimental values of the masses of the $Z_c(4025)$, $Y(3940)$ and $Y(4140)$, the $J=0,1,2$ denote the angular momentum of the molecular states, the (I) and (II) denote the $D^*\bar{D}^*$ and $D_s^*\bar{D}_s^*$ molecular states, respectively. The $D=8$ and $D=6$ denote the vacuum condensates are taken into account up to dimensions 8 and 6, respectively. }
\end{figure}

\begin{figure}
\centering
\includegraphics[totalheight=6cm,width=7cm]{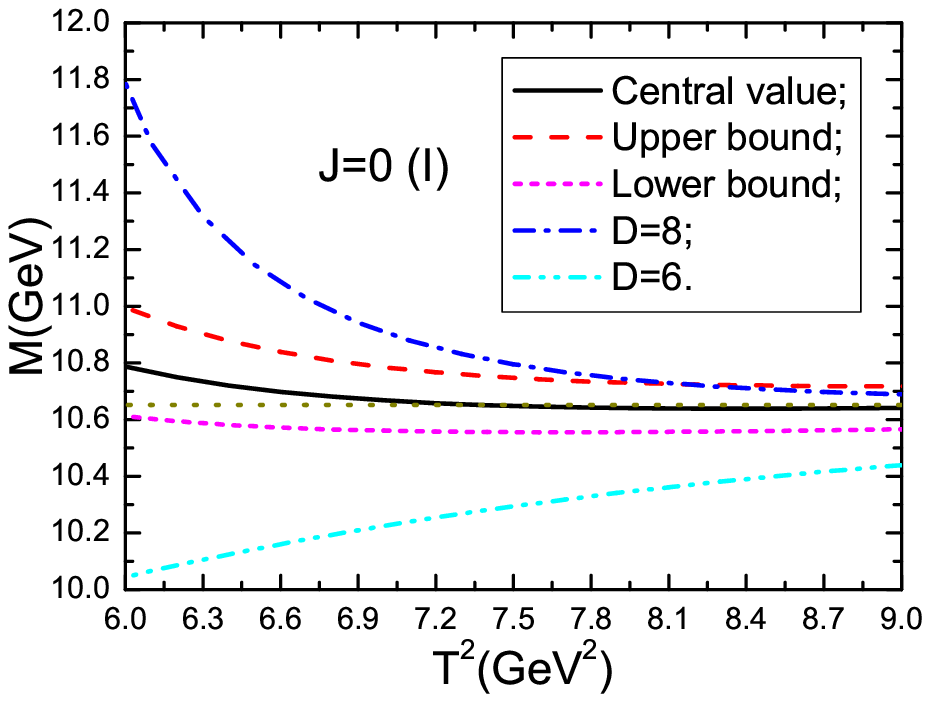}
\includegraphics[totalheight=6cm,width=7cm]{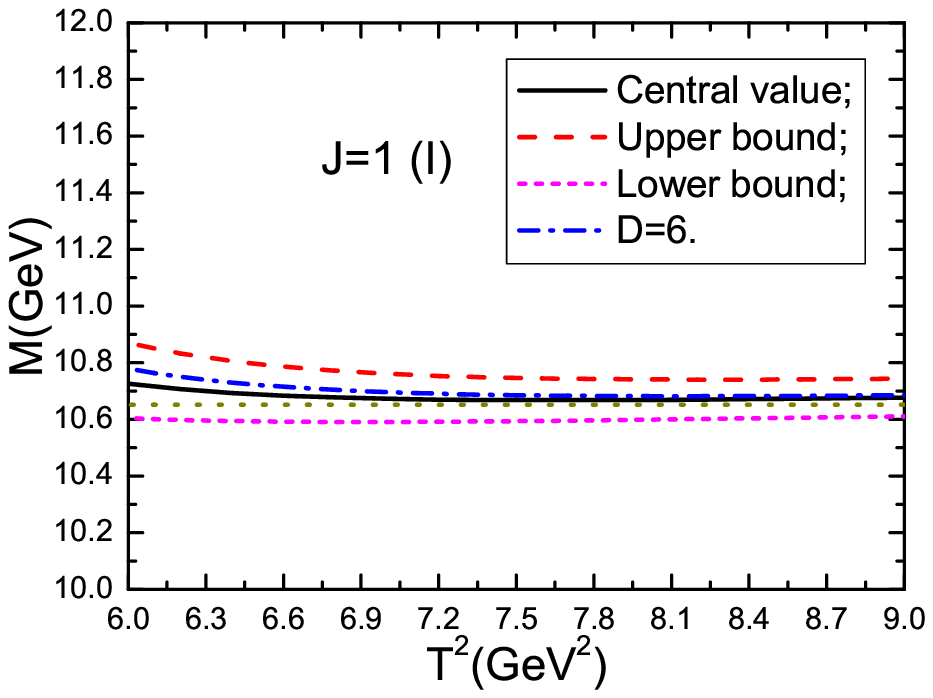}
\includegraphics[totalheight=6cm,width=7cm]{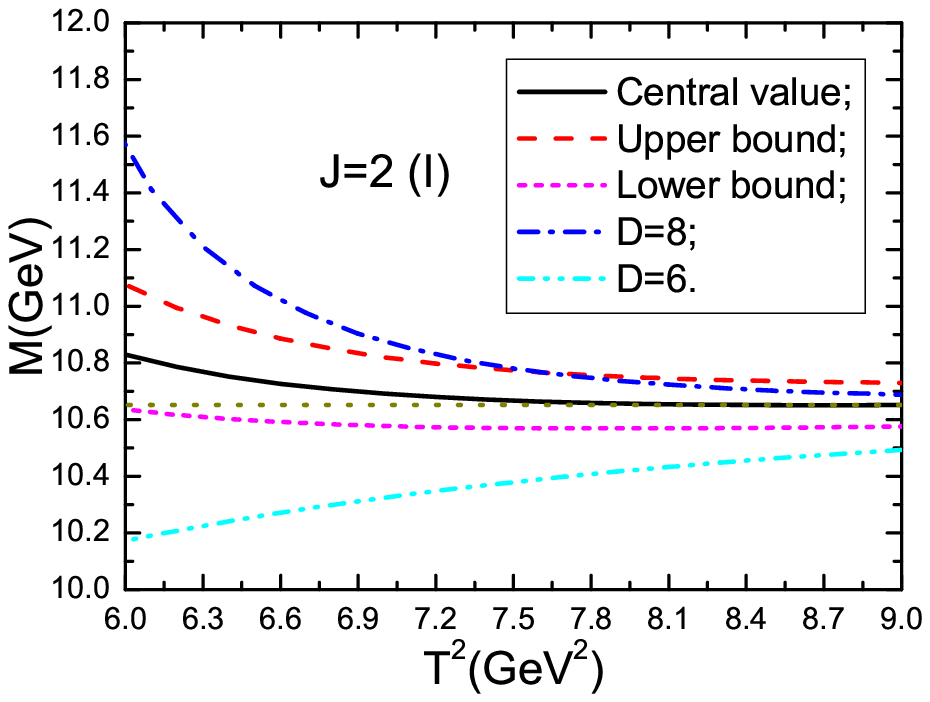}
\includegraphics[totalheight=6cm,width=7cm]{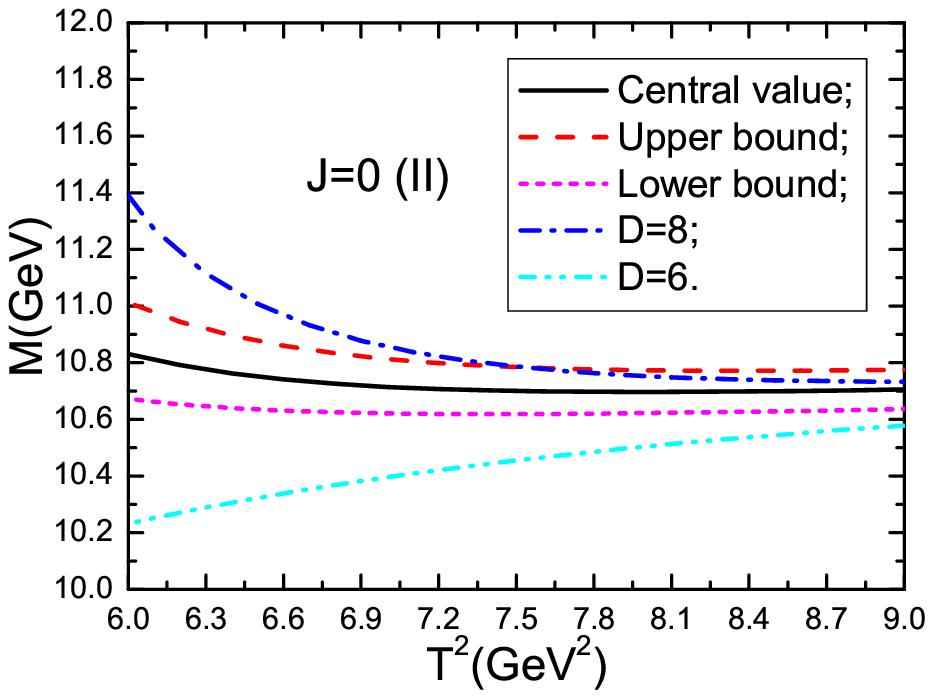}
\includegraphics[totalheight=6cm,width=7cm]{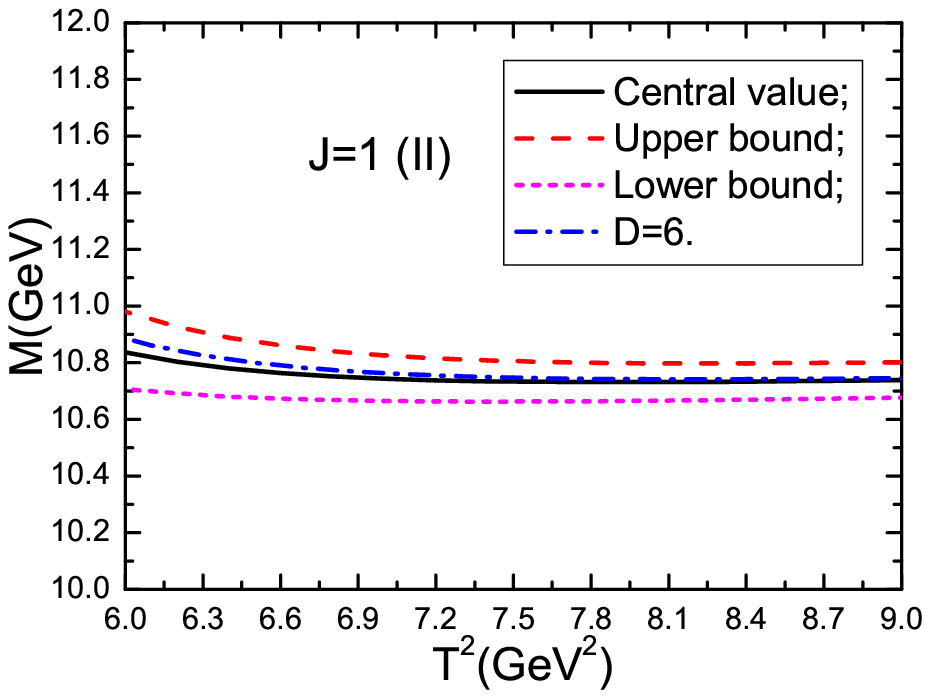}
\includegraphics[totalheight=6cm,width=7cm]{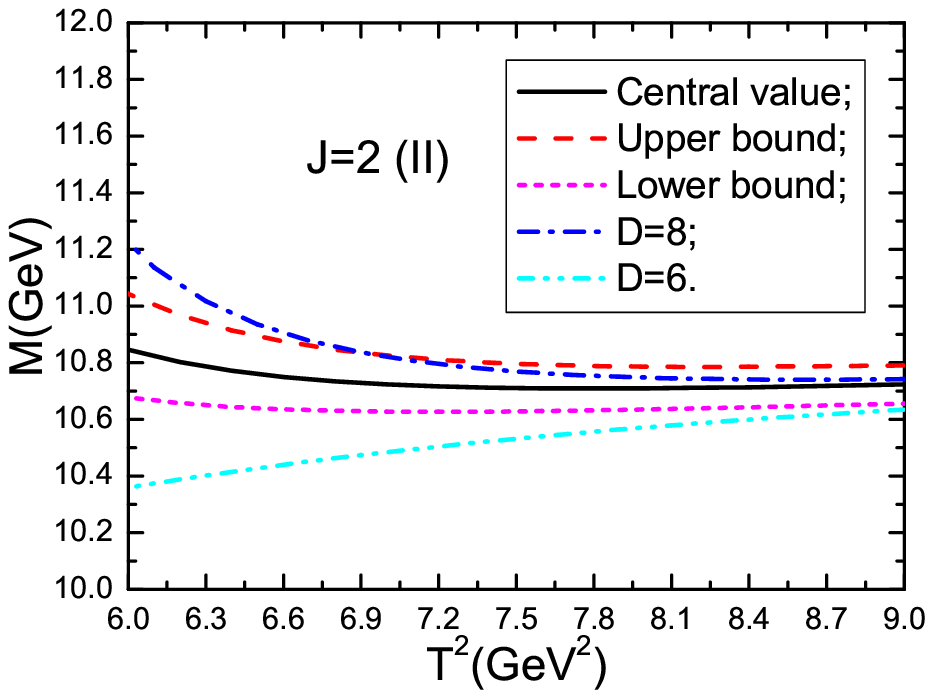}
  \caption{ The masses  with variations of the  Borel parameters $T^2$, where the horizontal lines denote the experimental value of the mass of the $Z_b(10650)$, the $J=0,1,2$ denote the angular momentum of the molecular states, the (I) and (II) denote the $B^*\bar{B}^*$ and $B_s^*\bar{B}_s^*$ molecular states, respectively. The $D=8$ and $D=6$ denote the vacuum condensates are taken into account up to dimensions 8 and 6, respectively. }
\end{figure}

We take into account all uncertainties of the input parameters, and obtain the values of the masses and pole residues of
 the   scalar, axial-vector and tensor molecular  states, which are  shown explicitly in Figs.4-5 and Table 1.

The uncertainties of the effective masses ${\mathbb{M}}_Q$ and energy scales $\mu$ have the correlation,
 \begin{eqnarray}
 4{\mathbb{M}}_Q\delta {\mathbb{M}}_Q &=&-\mu\delta\mu\, .
 \end{eqnarray}
 If we take the uncertainty $\delta\mu=0.3\,\rm{GeV}$, the induced uncertainties are  $\delta{\mathbb{M}}_c\approx 0.07\,\rm{GeV}$, $\delta{\mathbb{M}}_b\approx0.04\,\rm{GeV}$,  $\delta M_{Y/Z_b}\approx 100\,\rm{MeV}$, $\delta M_{Y/Z_c}\approx50\,\rm{MeV}$, $\delta M_{Y/Z}/M_{Y/Z}\approx 1\%$, $\delta \lambda_{Y/Z_c}/\lambda_{Y/Z_c}\approx 10\%$ and $\delta \lambda_{Y/Z_b}/\lambda_{Y/Z_b}\approx 20\%$, see Table 2.
 The uncertainties $\delta M_{Y/Z}/M_{Y/Z} \ll \delta \lambda_{Y/Z}/\lambda_{Y/Z}$, we obtain the hadronic masses $M_{Y/Z}$ through a ratio, see Eq.(46),
 the energy scale dependence of the hadronic masses $M_{Y/Z}$ originate from the numerator and denominator are canceled out with each other  efficiently, the predicted masses are robust.
 On the other hand, if we take  the uncertainties of the experimental values of the masses of the $Z_c(4020)$, $Z_c(4025)$, $Y(4140)$, $Z_b(10650)$ as the input parameters \cite{CDF0903,Belle1110,BES1308,BES1309}, the allowed uncertainties are  $|\delta\mu|\ll0.1\,\rm{GeV}$, $\delta{\mathbb{M}}_c\ll 0.03\,\rm{GeV}$, $\delta{\mathbb{M}}_b\ll 0.02\,\rm{GeV}$.
In Refs.\cite{Y4140-Nielsen,Y4140-Zhang,Tetraquark-Zb-QCDSR,DvDv-Nielsen}, the authors  study the $D^*\bar{D}^*$, $D_s^*\bar{D}_s^*$, $B^*\bar{B}^*$ molecular states by choosing the $\overline{MS}$ masses  $m_Q(m_Q)$ and the vacuum condensates $\langle\bar{q}q\rangle_{\mu=1\,\rm{GeV}}$, $\langle\bar{q}g_s\sigma Gq\rangle_{\mu=1\,\rm{GeV}}$, etc. In this article, we calculate the QCD spectral densities at a special energy scale $\mu$ consistently, the energy scales $\mu$ are determined by the parameters ${\mathbb{M}}_Q$, which have very small allowed uncertainties.
The correlation functions $\Pi(p)$ can be written as
\begin{eqnarray}
\Pi(p)&=&\int_{4m^2_Q(\mu)}^{s_0} ds \frac{\rho_{QCD}(s,\mu)}{s-p^2}+\int_{s_0}^\infty ds \frac{\rho_{QCD}(s,\mu)}{s-p^2} \, ,
\end{eqnarray}
through dispersion relation at the QCD side, and they are scale independent,
\begin{eqnarray}
\frac{d}{d\mu}\Pi(p)&=&0\, ,
\end{eqnarray}
which does not mean
\begin{eqnarray}
\frac{d}{d\mu}\int_{4m^2_Q(\mu)}^{s_0} ds \frac{\rho_{QCD}(s,\mu)}{s-p^2}\rightarrow 0 \, ,
\end{eqnarray}
 due to the following two reasons inherited from the QCD sum rules:\\
$\bullet$ Perturbative corrections are neglected, the higher dimensional vacuum condensates are factorized into lower dimensional ones therefore  the energy scale dependence of the higher dimensional vacuum condensates is modified;\\
$\bullet$ Truncations $s_0$ set in, the correlation between the threshold $4m^2_Q(\mu)$ and continuum threshold $s_0$ is unknown,  the quark-hadron duality is an assumption. \\
We cannot obtain energy scale independent QCD sum rules, but we have an energy scale formula to determine the energy scales consistently.

\begin{table}
\begin{center}
\begin{tabular}{|c|c|c|c|c|c|c|c|}\hline\hline
$J^{PC}$                       &$\mu(\rm{GeV})$  &$\delta{\mathbb{M}}_Q(\rm{GeV})$  &$\delta M_{Y/Z}(\rm{GeV})$       &$\delta\lambda_{Y/Z}/\lambda_{Y/Z}$ \\ \hline

$0^{++}$ ($c\bar{c}u\bar{d}$)  &$1.6\pm0.3$      &$\pm0.07$                         &${}^{+0.07}_{-0.05}$             &${}^{+9\%}_{-12\%}$ \\ \hline
$1^{+-}$ ($c\bar{c}u\bar{d}$)  &$1.7\pm0.3$      &$\pm0.07$                         &${}^{+0.06}_{-0.05}$             &${}^{+9\%}_{-12\%}$ \\ \hline
$2^{++}$ ($c\bar{c}u\bar{d}$)  &$1.6\pm0.3$      &$\pm0.07$                         &${}^{+0.08}_{-0.05}$             &${}^{+8\%}_{-11\%}$ \\ \hline

$0^{++}$ ($c\bar{c}s\bar{s}$)  &$1.8\pm0.3$      &$\pm0.07$                         &${}^{+0.05}_{-0.03}$             &${}^{+7\%}_{-9\%}$ \\ \hline
$1^{+-}$ ($c\bar{c}s\bar{s}$)  &$1.9\pm0.3$      &$\pm0.08$                         &${}^{+0.05}_{-0.03}$             &${}^{+6\%}_{-9\%}$ \\ \hline
$2^{++}$ ($c\bar{c}s\bar{s}$)  &$1.8\pm0.3$      &$\pm0.07$                         &${}^{+0.05}_{-0.03}$             &${}^{+6\%}_{-8\%}$ \\ \hline

$0^{++}$ ($b\bar{b}u\bar{d}$)  &$2.8\pm0.3$      &$\pm0.04$                         &${}^{+0.14}_{-0.10}$           &${}^{+18\%}_{-19\%}$ \\ \hline
$1^{+-}$ ($b\bar{b}u\bar{d}$)  &$2.9\pm0.3$      &$\pm0.04$                         &${}^{+0.12}_{-0.10}$           &${}^{+19\%}_{-20\%}$ \\ \hline
$2^{++}$ ($b\bar{b}u\bar{d}$)  &$2.8\pm0.3$      &$\pm0.04$                         &${}^{+0.13}_{-0.11}$           &${}^{+17\%}_{-19\%}$ \\ \hline

$0^{++}$ ($b\bar{b}s\bar{s}$)  &$2.9\pm0.3$      &$\pm0.04$                         &${}^{+0.12}_{-0.10}$           &${}^{+16\%}_{-18\%}$ \\ \hline
$1^{+-}$ ($b\bar{b}s\bar{s}$)  &$3.0\pm0.3$      &$\pm0.04$                         &${}^{+0.11}_{-0.09}$           &${}^{+18\%}_{-19\%}$ \\ \hline
$2^{++}$ ($b\bar{b}s\bar{s}$)  &$3.0\pm0.3$      &$\pm0.04$                         &${}^{+0.11}_{-0.08}$           &${}^{+15\%}_{-16\%}$ \\ \hline
 \hline
\end{tabular}
\end{center}
\caption{ The uncertainties originate from the uncertainty of the energy scale  $\delta\mu=0.3\,\rm{GeV}$.   }
\end{table}

The present predictions
$M^{J=2}_{D^*\bar{D}^*} =\left(4.01^{+0.10}_{-0.08}\right)\,\rm{GeV}$,
$M^{J=1}_{D^*\bar{D}^*} =\left(4.04^{+0.07}_{-0.08}\right)\,\rm{GeV}$,
$M^{J=0}_{D^*\bar{D}^*} =\left(4.01^{+0.09}_{-0.09}\right)\,\rm{GeV}$
 are consistent with the experimental values $M_{Z_c(4025)}=(4026.3\pm2.6\pm3.7)\,\rm{MeV}$, $M_{Z_c(4020)}=(4022.9\pm 0.8\pm 2.7)\,\rm{MeV}$ from the BESIII collaboration \cite{BES1308,BES1309}. More experimental data on the spin and parity are still needed to identify the $Z_c(4020)$ and $Z_c(4025)$ unambiguously.  In Ref.\cite{DvDv-Nielsen}, K. P. Khemchandani et al carry out the operator product expansion up to dimension-6 and obtain the values  $M^{J=2}_{D^*\bar{D}^*} = \left(3946 \pm 104 \right) \,\rm{MeV}$, $M^{J=1}_{D^*\bar{D}^*} = \left(3950 \pm 105\right) \,\rm{MeV}$, $ M^{J=0}_{D^*\bar{D}^*} = \left(3943 \pm 104\right) \,\rm{MeV}$. The central values are smaller than ours  about $50\,\rm{MeV}$.
 In calculations, we observe that the vacuum condensates of dimensions $7$, $8$, $10$ play an important role in determining  the Borel windows, and warrant platforms for
 the masses and pole residues. The conclusion survives in the QCD sum rules for the tetraquark states and molecular states consist of two heavy quarks and two light quarks. There appear terms of the orders $\mathcal{O}\left(\frac{1}{T^2}\right)$, $\mathcal{O}\left(\frac{1}{T^4}\right)$, $\mathcal{O}\left(\frac{1}{T^6}\right)$ in the QCD spectral densities, if we take into account the vacuum condensates whose dimensions are larger than 6 \cite{WangHuang-molecule,WangHuangTao1312,Wang1311,WangHuangTao}.
 The terms associate with $\frac{1}{T^2}$, $\frac{1}{T^4}$, $\frac{1}{T^6}$ in the QCD spectral densities  manifest themselves at small values of the Borel parameter $T^2$, we have to choose large values of the $T^2$ to warrant convergence of the operator product expansion and appearance of the Borel platforms. In the Borel windows, the higher dimension vacuum condensates  play a less important role.
In summary, the higher dimension vacuum condensates play an important role in determining the Borel windows therefore the ground state  masses and pole residues, so we should take them into account consistently.
  In Fig.4-5, we also plot the masses by taking into account the vacuum condensates up to dimension 6 and 8, respectively. From the figures, we can see that  neglecting the vacuum condensates of the dimensions 7, 8, 10 cannot lead to platforms  flat  enough so as to extract robust values.

  The present predictions
$M^{J=2}_{D_s^*\bar{D}_s^*} =\left(4.13^{+0.08}_{-0.08}\right)\,\rm{GeV}$,
$M^{J=1}_{D_s^*\bar{D}_s^*} =\left(4.16^{+0.05}_{-0.04}\right)\,\rm{GeV}$,
$M^{J=0}_{D_s^*\bar{D}_s^*} =\left(4.14^{+0.08}_{-0.08}\right)\,\rm{GeV}$
 are consistent with the experimental value $M_{Y(4140)}=(4143.0\pm2.9\pm1.2)\,\rm{MeV}$ from the CDF collaboration \cite{CDF0903}.
The CMS collaboration fitted  the peaking structure in the $J/\psi\phi$  mass spectrum to a $S$-wave relativistic Breit-Wigner line-shape with a statistical significance exceeding $5 \sigma$ \cite{CMS1309}. We can tentatively assign the $Y(4140)$ as the scalar $D_s^*\bar{D}_s^*$ molecular state, while there lack experimental candidates for the axial-vector and tensor $D_s^*\bar{D}_s^*$ molecular states. We can search for the axial-vector and tensor $D_s^*\bar{D}_s^*$ molecular states in the $J/\psi\phi$ mass spectrum and measure the angular correlation to determine the spin and parity.

 The present predictions
$M^{J=2}_{B^*\bar{B}^*} =\left(10.66^{+0.14}_{-0.09}\right)\,\rm{GeV}$,
$M^{J=1}_{B^*\bar{B}^*} =\left(10.67^{+0.09}_{-0.08}\right)\,\rm{GeV}$,
$M^{J=0}_{B^*\bar{B}^*} =\left(10.65^{+0.15}_{-0.09}\right)\,\rm{GeV}$
 are consistent with the experimental value  $M_{Z_b(10650)}=(10652.2\pm1.5)\,\rm{MeV}$ from the Belle collaboration \cite{Belle1110}, while the Belle data favors the
 $J^{PC}=1^{+-}$ assignment. We can tentatively assign the $Z_b(10650)$ as the axial-vector $B^*\bar{B}^*$ molecular state, while there lack experimental candidates for the scalar and tensor $B^*\bar{B}^*$ molecular states. We can search for the scalar and tensor $B^*\bar{B}^*$ molecular states in the $\Upsilon\varphi$  mass spectrum   and measure the angular correlations  to determine the spin and parity.

 There also lack experimental candidates for the $B_s^*\bar{B}_s^*$ molecular states, we can search for them in the $\Upsilon\phi$  mass spectrum   and measure the angular correlations to determine the spin and parity.

In Refs.\cite{WangHuangTao1312,Wang1311}, we resort to  the same routine  to study the  heavy tetraquark states, the predicted masses favor assigning the $Z_c(4020)$ and  $Z_c(4025)$ as the $1^{+-}$ or $2^{++}$ tetraquark states, the $Z_b(10650)$ as the $1^{+-}$ tetraquark state.  A hadron cannot be identified  unambiguously by  the mass alone \cite{Tetraquark-Zb-QCDSR}, so it is interesting to explore possible assignments in the scenario of molecular states. The predicted masses of the heavy molecular states also favor  assigning
the $Z_c(4020)$ and  $Z_c(4025)$ as the $1^{+-}$ or $2^{++}$ molecular states, the $Z_b(10650)$ as the $1^{+-}$ molecular state. The $Z_c(4020)$, $Z_c(4025)$, $Z_b(10650)$ maybe have both tetraquark and molecule components, which should be interpolated by the tetraquark-type currents and molecule-type  currents, respectively. In the present work and Refs.\cite{WangHuang-molecule,WangHuangTao1312,Wang1311,WangHuangTao}, we obtain the pole residues (or the current-hadron coupling constants), which can be taken as basic input parameters to study the strong decays of the heavy tetraquark states or molecular states with the three-point QCD sum rules. Then we obtain more  knowledge to identify the $Z_c(4020)$, $Z_c(4025)$, $Z_b(10650)$.
In the scenario of meta-stable Feshbach resonances, the  $Z_c(4025)$ and $Z_b(10650)$ are taken as the
 $h_c({\rm 2P})\pi-D^*\bar{D}^*$ and $\chi_{b1}\rho-B^*\bar{B}^*$ hadrocharmonium-molecule
 mixed states, respectively, where the  $\chi_{b1}\rho$ is a P-wave system \cite{Italy1311}. The hadrocharmonium system admits bound states giving rise to a
discrete spectrum of levels, a resonance occurs if one of such levels falls close to some
open-charm (open-bottom) threshold, as  the coupling between channels leads  to an attractive interaction and favors the formation of a meta-stable Feshbach
resonance. We can borrow some ideas from the meta-stable Feshbach resonances, the couplings between the tetraquark states and molecular states leads to an attractive interaction and favors the formation of the $Z_c(4020)$, $Z_c(4025)$, $Z_b(10650)$, as they couple potentially both to the tetraquark type and molecule type currents.

\section{Conclusion}
In this article, we calculate the contributions of the vacuum condensates up to dimension-10 and discard the perturbative corrections in the operator product expansion, and  study the  $J^{PC}=0^{++}$, $1^{+-}$ and $2^{++}$  $D^*\bar{D}^*$, $D_s^*\bar{D}_s^*$,  $B^*\bar{B}^*$, $B_s^*\bar{B}_s^*$ molecular states   in details with the QCD sum rules. In calculations,  we use the  formula $\mu=\sqrt{M^2_{X/Y/Z}-(2{\mathbb{M}}_Q)^2}$ suggested in our previous work to determine  the energy scales of the QCD spectral densities.  The present predictions favor assigning the $Z_c(4020)$ and $Z_c(4025)$  as the $J^{PC}=0^{++}$, $1^{+-}$ or $2^{++}$    $D^*\bar{D}^*$ molecular states, the $Y(4140)$ as the $J^{PC}=0^{++}$ $D^*_s{D}_s^*$ molecular state, the $Z_b(10650)$  as the $J^{PC}=1^{+-}$    $B^*\bar{B}^*$ molecular state, and  disfavor assigning the $Y(3940)$  as the ($J^{PC}=0^{++}$) molecular state.  The present  predictions can be confronted with the experimental data in the futures at the BESIII, LHCb and Belle-II. The  pole residues can be taken as   basic input parameters to study relevant processes of the $J^{PC}=0^{++}$, $1^{+-}$ and $2^{++}$ molecular  states with the three-point QCD sum rules.

\section*{Acknowledgements}
This  work is supported by National Natural Science Foundation,
Grant Number 11375063, the Fundamental Research Funds for the
Central Universities, and Natural Science Foundation of Hebei province, Grant Number A2014502017.


\begin{thebibliography}{99}


\bibitem{Belle2004} S. K. Choi  et al,  Phys. Rev. Lett. {\bf 94} (2005) 182002.

\bibitem{BaBar2007} B. Aubert et al, Phys. Rev. Lett. {\bf 101} (2008) 082001.

\bibitem{Belle2010} S. Uehara  et al,  Phys. Rev. Lett. {\bf 104} (2010) 092001.

\bibitem{PDG}  J. Beringer et al, Phys. Rev. {\bf D86} (2012) 010001.


\bibitem{CDF0903} T. Aaltonen et al,  Phys. Rev. Lett. {\bf 102} (2009) 242002.

 \bibitem{Belle0912} C. P. Shen et al,  Phys. Rev. Lett. {\bf 104} (2010) 112004.

\bibitem{LHCb1202} R. Aaij  et al,  Phys. Rev. {\bf D85} (2012) 091103.

\bibitem{CMS1309} S. Chatrchyan et al, arXiv:1309.6920.


\bibitem{D0-1309}  V. M. Abazov,  Phys. Rev. {\bf D89} (2014) 012004.


\bibitem{Y4140-molecule} X. Liu  and S. L. Zhu, Phys. Rev. {\bf D80} (2009) 017502;
 T. Branz, T. Gutsche and V. E. Lyubovitskij, Phys. Rev. {\bf D80} (2009) 054019;
  G. J. Ding, Eur. Phys. J. {\bf C64} (2009) 297;
 R. Molina and E. Oset, Phys. Rev. {\bf D80} (2009) 114013;
 X. Liu, Z. G. Luo and S. L. Zhu, Phys. Lett. {\bf B699} (2011) 341.



\bibitem{Y4140-Wang} Z. G. Wang, Eur. Phys. J. {\bf C63} (2009) 115;
Z. G. Wang, Z. C. Liu and X. H. Zhang,  Eur. Phys. J. {\bf C64} (2009) 373.


\bibitem{Y4140-Nielsen} R. M. Albuquerque, M. E. Bracco and M. Nielsen, Phys. Lett. {\bf B678} (2009) 186.

\bibitem{Y4140-Zhang} J. R. Zhang and M. Q. Huang, J. Phys. {\bf G37} (2010) 025005.


\bibitem{Y4140-hybrid}   N. Mahajan, Phys. Lett. {\bf B679} (2009) 228;


\bibitem{Y4140-tetraquark} F. Stancu, J. Phys. {\bf G37} (2010) 075017.


\bibitem{Belle1105} I. Adachi et al,  arXiv:1105.4583.

\bibitem{Belle1110}   A. Bondar   et al,  Phys. Rev. Lett. {\bf 108} (2012) 122001.


\bibitem{Belle1308} P. Krokovny  et al, Phys. Rev. {\bf D88} (2013) 052016.

\bibitem{Molecule-Zb} A. E. Bondar, A. Garmash, A. I. Milstein, R. Mizuk and M. B. Voloshin, Phys. Rev. {\bf D84} (2011) 054010;
J. R. Zhang, M. Zhong and M. Q. Huang,  Phys. Lett. {\bf B704} (2011) 312;
M. B. Voloshin,  Phys. Rev. {\bf D84} (2011) 031502;
J. Nieves and M. Pavon Valderrama, Phys. Rev. {\bf D84} (2011) 056015;
Z. F. Sun, J. He, X. Liu, Z. G. Luo and S. L. Zhu, Phys. Rev. {\bf D84} (2011) 054002;
M. Cleven, F. K. Guo, C. Hanhart and Ulf-G. Meissner, Eur. Phys. J. {\bf A47} (2011) 120;
 T. Mehen and J. W. Powell, Phys. Rev. {\bf D84} (2011) 114013;
Y. Yang , J. Ping, C. Deng and H. S. Zong, J. Phys. {\bf G39} (2012) 105001;
S. Ohkoda, Y. Yamaguchi, S. Yasui, K. Sudoh and A. Hosaka, Phys. Rev. {\bf D86} (2012) 014004;
 H. W. Ke, X. Q. Li, Y. L. Shi, G. L. Wang and X. H. Yuan, JHEP {\bf 1204} (2012) 056;
Y. Dong, A. Faessler, T. Gutsche and V. E. Lyubovitskij, J. Phys. {\bf G40} (2013) 015002;
 M. B. Voloshin,  Phys. Rev. {\bf D87} (2013) 074011.

\bibitem{WangHuang-molecule} Z. G. Wang and T. Huang, Eur. Phys. J. {\bf C74} (2014) 2891.


\bibitem{Tetraquark-Zb} A. Ali and C. Hambrock and W. Wang,  Phys. Rev. {\bf D85} (2012) 054011.


\bibitem{Tetraquark-Zb-QCDSR} C. Y. Cui, Y. L. Liu and M. Q. Huang, Phys. Rev. {\bf D85} (2012) 074014.

\bibitem{WangHuangTao1312} Z. G. Wang and T. Huang,  arXiv:1312.2652.





\bibitem{Cusp-Zb} D. V. Bugg, Europhys. Lett. {\bf 96} (2011) 11002.


\bibitem{Rescatter-Zb}  D. Y. Chen, X. Liu and S. L. Zhu, Phys. Rev. {\bf D84} (2011) 074016;
G. Li, F. l. Shao, C. W. Zhao and Q. Zhao,  Phys. Rev. {\bf D87} (2013) 034020.


\bibitem{BES1308}   M. Ablikim  et al, Phys. Rev. Lett. {\bf 112} (2014) 132001.

\bibitem{BES1309}  M. Ablikim  et al, Phys. Rev. Lett. {\bf 111} (2013) 242001.



\bibitem{Molecule} F. K. Guo, C. Hidalgo-Duque, J. Nieves and M. P. Valderrama, Phys. Rev. {\bf D88} (2013) 054007;
J. He, X. Liu, Z. F. Sun and S. L. Zhu, Eur. Phys. J. {\bf C73} (2013) 2635;
A. Martinez Torres, K. P. Khemchandani, F. S. Navarra, M. Nielsen and E. Oset, Phys. Rev. {\bf D89} (2014) 014025.

\bibitem{Cui-DvDv} C. Y. Cui, Y. L. Liu and M. Q. Huang, Eur. Phys. J. {\bf C73} (2013) 2661.


\bibitem{Chen-Zhu} W. Chen, T. G. Steele, M. L. Du and S. L. Zhu, Eur. Phys. J. {\bf C74} (2014) 2773.


\bibitem{DvDv-Nielsen} K. P. Khemchandani, A. Martinez Torres, M. Nielsen and F. S. Navarra, Phys. Rev. {\bf D89} (2014) 014029.

\bibitem{Rescatter}  G. Li, Eur. Phys. J. {\bf C73} (2013) 2621;
X. Wang, Y. Sun, D. Y. Chen, X. Liu and T. Matsuki, Eur. Phys. J. {\bf C74} (2014) 2761.

\bibitem{Tetraquark-Qiao} C. F. Qiao and L. Tang,  Eur. Phys. J. {\bf C74} (2014) 2810.

 \bibitem{Wang1311} Z. G. Wang,  Eur. Phys. J. {\bf C74} (2014) 2874; Z. G. Wang, arXiv:1312.1537.

 \bibitem{WangHuangTao} Z. G. Wang and T. Huang,  Phys. Rev. {\bf D89} (2014) 054019.



\bibitem{Swanson2006} E. S. Swanson, Phys. Rept. {\bf 429} (2006) 243;
S. Godfrey and S. L. Olsen, Ann. Rev. Nucl. Part. Sci. {\bf 58} (2008) 51;
M. B. Voloshin, Prog. Part. Nucl. Phys. {\bf 61} (2008) 455;
N. Drenska, R. Faccini, F. Piccinini, A. Polosa, F. Renga and C. Sabelli, Riv. Nuovo Cim. {\bf 033} (2010) 633;
 N. Brambilla et al,  Eur. Phys. J. {\bf C71} (2011) 1534.


\bibitem{SVZ79}  M. A. Shifman, A. I. Vainshtein and V. I. Zakharov, Nucl. Phys. {\bf B147} (1979) 385.

\bibitem{Reinders85} L. J. Reinders, H. Rubinstein and S. Yazaki, Phys. Rept. {\bf 127} (1985) 1.

\bibitem{WangHcHb} Z. G. Wang, Eur. Phys. J. {\bf C73} (2013) 2533.

\bibitem{external} W. Hubschmid and S. Mallik, Nucl. Phys. {\bf B207} (1982) 29;
V. A. Novikov, M. A. Shifman, A. I. Vainshtein and V. I. Zakharov, Fortsch. Phys. {\bf 32} (1984) 585.


\bibitem{Ioffe2005} P. Colangelo and A. Khodjamirian, hep-ph/0010175;
B. L. Ioffe, Prog. Part. Nucl. Phys. {\bf 56} (2006) 232.


\bibitem{Italy1311} M. Papinutto, F. Piccinini, A. Pilloni, A. D. Polosa and N. Tantalo, arXiv:1311.7374.

\end{thebibliography}
\end{document}